   \newcommand{\be}[0]{\begin{equation}}
   \newcommand{\ee}[0]{\end{equation}}
   \newcommand{\ba}[0]{\begin{eqnarray}}
   \newcommand{\ea}[0]{\end{eqnarray}}
	\newcommand{\zrow}[5]{{#1}&{#2}&{#3}&{#4}&{#5}\\}
	
	\newcommand{\MS}[0]{\overline{{\rm MS}} }
\documentstyle[12pt,epsfig]{article}

\begin{document}
\topmargin=-0.5in
\headheight=0.6in
\headsep=0in
\footheight=0.5in
\footskip=0.5in
\Large
\hfill\vbox{\hbox{DTP/95/36}
            \hbox{hep-ph/9505224}
            \hbox{May 1995}
            \hbox{Revised June 1995}}
\nopagebreak

\vspace{0.75cm}
\begin{center}
\LARGE
{\bf All-Orders Renormalon Resummations for some QCD Observables}
\vspace{0.6cm}
\Large

C.N.Lovett-Turner and C.J.Maxwell

\vspace{0.4cm}
\large
\begin{em}
Centre for Particle Theory, University of Durham\\
South Road, Durham, DH1 3LE, England
\end{em}

\vspace{1.7cm}

\end{center}
\normalsize
\vspace{0.45cm}

Exact large-$N_{f}$ results for the QCD Adler $D$-function and
Deep Inelastic Scattering sum rules are used to resum to all orders
the portion of QCD perturbative coefficients containing the highest
power of $b$=$\frac{1}{6}(11N$--$2N_{f})$, for SU($N$) QCD with
$N_{f}$ quark flavours. These terms correspond to renormalon
singularities in the Borel plane and are expected asymptotically to
dominate the coefficients to all orders in the $1/N_{f}$ expansion.
Remarkably, we note that this is already apparent in comparisons with
the exact next-to-leading order (NLO) and next-to-NLO (NNLO)
perturbative coefficients. The ultra-violet ($UV$) and infra-red
($IR$) renormalon singularities in the Borel transform are isolated
and the Borel sum (principal value regulated for $IR$) performed.
Resummed results are also obtained for the Minkowski quantities
related to the $D$-function, the $e^{+}e^{-}$ $R$-ratio and the
analogous $\tau$-lepton decay ratio, $R_{\tau}$. The renormalization
scheme dependence of these partial resummations is discussed and they
are compared with the results from other groups [1--3] and with exact
fixed order perturbation theory at NNLO. Prospects for improving the
resummation by including more exact details of the Borel transform
are considered.
\newpage

\section{Introduction}

In several recent papers [1--3] the possibility of resummation to all
orders of the part of perturbative corrections contributed by QCD
renormalons has been explored.

The QCD perturbative corrections to some generic QCD Green's function
or current correlator (to the Adler $D$-function of QCD vacuum
polarization for instance) can be written:
\be
D=a+d_{1}a^{2}+d_{2}a^{3}+\cdots+d_{k}a^{k+1}+\cdots\;,
\ee
where
$a\equiv\alpha_{s}/\pi$ is the renormalization group (RG) improved
coupling; and the perturbative coefficients $d_{k}$ can themselves be
written as polynomials of degree $k$ in the number of quark flavours,
$N_{f}$; we shall assume massless quarks.
\be
d_{k}=d_{k}^{[k]}N_{f}^{k}+d_{k}^{[k-1]}N_{f}^{k-1}+
\cdots+d_{k}^{[0]}\;.
\ee
The ``$N_{f}$-expansion'' coefficients,
$d_{k}^{[k-r]}$, will consist of sums of multinomials in the adjoint
and
fundamental Casimirs, $C_{A}$=$N$, $C_{F}$=$(N^{2}$--$1)/2N$, of
SU($N$)
QCD;
and will have the structure $C_{A}^{k-r-s}C_{F}^{s}$. The terms in
this ``$N_{f}$-expansion'' will correspond to Feynman diagrams with
differing numbers of vacuum polarization loops. By explicit
evaluation
of diagrams with chains of such loops inserted, it has been possible
to obtain the leading $d_{k}^{[k]}$ coefficient exactly to all orders
for the Adler $D$-function [4--6] (and hence its Minkowski
continuations,
the $e^{+}e^{-}$ QCD $R$-ratio and the $\tau$-decay ratio,
$R_{\tau}$);
the Gross Llewellyn-Smith (GLS) sum rule corrections \cite{broadkat};
and heavy quark
decay widths and pole masses \cite{benbraun2}. A general procedure
enabling
$d_{k}^{[k]}$ to be obtained from knowledge of the one-loop correction
with a fictitious gluon mass has been developed
\cite{benbraun1,ballbenbraun}.

In a recent paper \cite{us} we pointed out that the large order
behaviour of
perturbative coefficients is most transparently discussed in terms of
an expansion of the perturbative coefficients in powers of
$b$=$(11C_{A}$--$2N_{f})/6$, the first QCD beta-function coefficient:
\be
d_{k}=d_{k}^{(k)}b^{k}+d_{k}^{(k-1)}b^{k-1}+\cdots+d_{k}^{(0)}\;.
\ee
This ``$b$-expansion'' is uniquely obtained by substituting
$N_{f}$=$(\frac{11}{2}C_{A}-3b)$ in equation (2).
$d_{k}^{[k]}$=$(-1/3)^{k}d_{k}^{(k)}$ and so exact knowledge of the
leading-$N_{f}$ $d_{k}^{[k]}$ to all orders implies exact knowledge
of the $d_{k}^{(k)}$.

In QCD one expects the large-order growth of perturbative coefficients
to be driven by Borel plane singularities at $z$=$z_{\ell}$=$2\ell/b$,
with $\ell$=$\pm1,\pm2,\pm3,\ldots$. The singularities on the negative
real axis are the so-called ultra-violet renormalons, $UV_{\ell}$,
and those on the positive real axis are the infra-red renormalons,
$IR_{\ell}$. These singularities result in the large-order behaviour
of the coefficients $d_{k}\sim b^{k}k!$. Indeed we showed in
reference \cite{us} that, given a set of renormalon singularities at
the
expected positions in the Borel plane, the leading terms in the
$b$-expansion, $d_{k}^{(k)}b^{k}$, should, if expanded in powers of
$N_{f}$, asymptotically reproduce the $d_{k}^{[k-r]}$ coefficients of
equation (2) up to $O(1/k)$ accuracy. We conversely checked that the
exact $d_{k}^{(k)}b^{k}$ results corresponded to a set of renormalon
singularities at the expected positions. We shall demonstrate in
section 2 of the present paper that, for the Adler $D$-function and
the GLS sum rule, the $N_{f}$-expansion coefficients obtained by
expanding $d_{1}^{(1)}b$ and $d_{2}^{(2)}b^{2}$ are in good (10--20\%
level) agreement with those of the exact $O(a^{3})$
next-to-next-to-leading order (NNLO) perturbative calculations for
these quantities, so that the anticipated asymptotic dominance of the
leading-$b$ term is already apparent in low orders.

Given the dominance of the leading-$b$ terms, an obvious proposal is
to sum them to all orders. That is to split $D$ into two components:
\be
D=D^{(L)}+D^{(NL)}
\ee
where `$L$' and `$NL$' superscripts refer to leading and non-leading
terms in the $b$-expansion.
\be
D^{(L)}\equiv\sum_{k=1}^{\infty}d_{k}^{(k)}b^{k}a^{k+1}
\ee
and
\be
D^{(NL)}\equiv\sum_{k=1}^{\infty}a^{k+1}\sum_{\ell=0}^
{k-1}d_{k}^{(\ell)}b^{\ell}\;.
\ee
The summation of terms can be achieved by using the Borel sum. The
Borel integral can itself be split into two components and is well
defined for the $UV_{\ell}$ singularities on the negative axis, which
contribute poles to the Borel transform of $D^{(L)}$. The integral
can be performed
explicitly in terms of exponential integral functions and other
elementary functions. The piece of the Borel integral for $D^{(L)}$
involving the $IR_{\ell}$ singularities on the positive real axis is
formally divergent; but a principal value or other prescription can be
used to go around the poles. The specification of this prescription is
intimately linked to the procedure needed to combine the
non-perturbative vacuum condensates in the operator product expansion
(OPE) with the perturbation theory in order to arrive at a
well-defined result
for $D$ \cite{grun}.

In recent papers by Neubert \cite{neubert} and by Ball, Beneke and
Braun \cite{benbraun1,ballbenbraun} a
summation of the leading-$b$ terms has also been considered. In these
papers it has been motivated as a generalisation of the BLM scale
fixing prescription \cite{blm} and termed ``na\"{\i}ve
non-abelianization''
\cite{broadgroz}. The Neubert procedure uses weighted integrals over
a running
coupling. For the Euclidean Adler $D$-function this representation is
equivalent to splitting the Borel integral into ultra-violet
renormalon and infra-red renormalon singularities and principal value
regulating the latter. When one continues to Minkowski space to
obtain the $e^{+}e^{-}$ $R$-ratio and the $\tau$-decay ratio,
$R_{\tau}$,
there are several inequivalent ways to perform the continuation of
the running coupling representation; and hence apparent additional
non-perturbative ambiguities are claimed.
In reference \cite{ballbenbraun} the resummation is defined by using
the principal value regulated Borel integral, as we shall do. They
concentrate on $R_{\tau}$ and heavy quark pole masses. We agree with
reference \cite{ballbenbraun} that only a
consideration of the singularities in the Borel integral provides a
satisfactory way of combining perturbative effects with
non-perturbative condensates, along the lines discussed in reference
\cite{grun};
and that the extra uncertainties claimed in reference \cite{neubert}
are spurious.

Our intention in this paper is to focus on the Adler $D$-function,
the $e^{+}e^{-}$ $R$-ratio, $R_{\tau}$ and the GLS sum rule (the
latter was not considered in references [1--3]). For all of these
quantities there exist exact NNLO fixed order perturbative
calculations and our interest is in comparing the leading-$b$
resummation with these exact fixed order results. The large-$b$
results provide partial information about the Borel transform and the
question is how this can best be utilised. We discuss the
renormalization scheme (RS) dependence of the split between $D^{(L)}$
and $D^{(NL)}$ in equation (4), the relative contribution of
$D^{(L)}$ being RS-dependent. This RS uncertainty needs to be kept in
mind and is carefully discussed.

The organisation of the paper is as follows. In section 2 we shall
discuss the $b$-expansion for the Adler $D$-function and the GLS sum
rule and will consider the extent to which the dominance of the
leading-$b$ term is RS-dependent. In section 3 the exact large-$b$
results are used to determine partially the Borel transforms for these
quantities; and, having split them into $UV$ and $IR$ renormalon
pieces, a resummation along the lines discussed above is performed.

In section 4 the RS dependence of the resummed results is considered,
the numerical results are presented and a comparison with the
results of references \cite{ballbenbraun,neubert} and with the exact NNLO
perturbative results is made.
A discussion
of the uncertainties and the prospects for improving the resummation
by including more of the exact structure of the Borel transform is
then undertaken. Section~5 contains overall conclusions.

\section{The $b$-expansions for $\tilde{D}$ and $\tilde{K}$}

We begin by defining the Adler $D$-function and the GLS sum rule, the
quantities with which we shall be concerned in this paper.

$D(Q^{2})$ is related to the vacuum polarization function, the
correlator, $\Pi(Q^{2})$, of two vector currents in the Euclidean
region,
\be
(q_{\mu}q_{\nu}-g_{\mu\nu}q^{2})\Pi(Q^{2})=16\pi^{2}i
\int\mbox{d}^{4}x\,e^{i{\bf q.x}}\langle 0|T\{J_{\mu}(x)
J_{\nu}(0)\}|0\rangle
\;,
\ee
by
\be
D(Q^{2})=-\frac{3}{4}Q^{2}\frac{\mbox{d}}{\mbox{d}Q^{2}}\Pi(Q^{2})\;,
\ee
with $Q^{2}$=$-q^{2}>0$.

In perturbation theory one has
\be
D(Q^{2})=d(R)\sum_{f}Q_{f}^{2}\left(1+\frac{3}{4}C_{F}\tilde{D}\right)
+\left(\sum_{f}Q_{f} \right)^{2}\tilde{\tilde{D}}
\;.
\ee
Here $d(R)$ is the dimension of the quark representation of the
colour
group, $d(R)$=$N$ for SU($N$) QCD, and $Q_{f}$ denotes the quark
charges, summed over the accessible flavours at a given energy.
$\tilde{\tilde{D}}$ denotes corrections of the ``light-by-light''
type which first enter at $O(a^{3})$ and will be subleading in
$N_{f}$;
they are, it is to be hoped, small. $\tilde{D}$ represents the QCD
corrections to the zeroth order parton model result and has the form
\be
\tilde{D}=a+d_{1}a^{2}+d_{2}a^{3}+\cdots+d_{k}a^{k+1}+\cdots\;.
\ee

We shall next consider two QCD deep inelastic scattering sum rules.
The first is the polarized Bjorken sum rule (PBjSR):
\ba
K_{PBj}&\equiv&\int_{0}^{1}g_{1}^{\mbox{\scriptsize
ep-en \normalsize}}
(x,Q^{2})\mbox{d}x
\nonumber\\
&=&\frac{1}{3}\left|\frac{g_{A}}{g_{V}}\right|\left(1-
\frac{3}{4}C_{F}\tilde{K}\right)
\;.
\ea
Here $\tilde{K}$ denotes the perturbative corrections to the zeroth
order parton model sum rule,
\be
\tilde{K}=a+K_{1}a^{2}+K_{2}a^{3}+\cdots+K_{k}a^{k+1}+\cdots\;.
\ee
We can also consider the GLS sum rule,
\ba
K_{GLS}&\equiv&\frac{1}{6}\int_{0}^{1}F_{3}^{\mbox{\scriptsize
$\overline{\nu}$p+$\nu$p \normalsize}}
(x,Q^{2})\mbox{d}x
\nonumber\\
&=&\left(1-\frac{3}{4}C_{F}\tilde{K}+\tilde{\tilde{K}}\right)
\;.
\ea
The perturbative corrections, $\tilde{K}$, are the same as for the
PBjSR;
but there are additional corrections of ``light-by-light'' type,
$\tilde{\tilde{K}}$, analogous to $\tilde{\tilde{D}}$ of equation (9).
These will similarly enter at $O(a^{3})$ and be subleading in $N_{f}$
and we shall assume once again that they are small.

For both $\tilde{D}$ and $\tilde{K}$ the first two perturbative
coefficients, $d_{1}$, $d_{2}$ and $K_{1}$, $K_{2}$, are known from
the
exact perturbative calculations [12--15]. We shall assume
$\MS$ renormalization with renormalization scale
$\mu$=$Q$ for the present; but will later discuss RS dependence more
generally.

Writing $d_{1}$ and $d_{2}$ expanded in $N_{f}$ as in equation (2),
the exact calculations give:
\ba
d_{1}&=&\left(-\frac{11}{12}+\frac{2}{3}\zeta_{3}\right)N_{f}
+C_{A}\left(\frac{41}{8}-\frac{11}{3}\zeta_{3}\right)
-\frac{1}{8}C_{F} \;,\\
d_{2}&=&\left(\frac{151}{162}-\frac{19}{27}\zeta_{3}\right)N_{f}^{2}
+C_{A}\left(-\frac{970}{81}+\frac{224}{27}\zeta_{3}+\frac{5}{9}
\zeta_{5}\right)N_{f}\nonumber\\
& &+C_{F}\left(-\frac{29}{96}+\frac{19}{6}\zeta_{3}-\frac{10}{3}
\zeta_{5}\right)N_{f}
+C_{A}^{2}\left(\frac{90445}{2592}-\frac{2737}{108}\zeta_{3}
-\frac{55}{18}\zeta_{5}\right)\nonumber\\
& &+C_{A}C_{F}\left(-\frac{127}{48}-\frac{143}{12}\zeta_{3}
+\frac{55}{3}\zeta_{5}\right)+C_{F}^{2}\left(-\frac{23}{32}\right)\;.
\ea
Here $\zeta_{3}$ and $\zeta_{5}$ are Riemann $\zeta$-functions. For
later
comparisons it will be useful to write these results numerically for
SU($N$) QCD.
\ba
d_{1}&=&-.115N_{f}+\left(.655N+\frac{.063}{N}\right) \;,\\
d_{2}&=&.086N_{f}^{2}+N_{f}\left(-1.40N-\frac{.024}{N}\right)
+\left(2.10N^{2}-.661-\frac{.180}{N^{2}}\right)
\;.
\ea
Expanding $d_{1}$ and $d_{2}$ in $b$ as in equation (3) gives
\ba
d_{1} &=& \left( \frac{11}{4}-2\zeta_{3}\right) b+\frac{C_{A}}{12}-
\frac{C_{F}}
{8}\;,\\
d_{2} &=& \left( \frac{151}{18}-\frac{19}{3}\zeta_{3}\right)b^{2}
+C_{A}\left(\frac{31}{6}-\frac{5}{3}\zeta_{3}-\frac{5}{3}
      \zeta_{5}
      \right) b\nonumber\\
      & & +C_{F}\left(\frac{29}{32}-\frac{19}{2}\zeta_{3}+10\zeta_{5}
      \right)b
      +C_{A}^{2}\left(-\frac{799}{288}-\zeta_{3}\right)\nonumber\\
      & & +C_{A}C_{F}\left(-\frac{827}{192}+\frac{11}{2}\zeta_{3}
      \right)+C_{F}^{2}\left(-\frac{23}{32}\right)\;.
\ea
As observed in reference \cite{us}, the $b$-expansion exhibits certain
simplifications relative to that in $N_{f}$. In particular the
$\zeta_{3}$, present in all orders of the $N_{f}$-expansion for
$d_{1}$,
is present only in the leading term in the $b$-expansion. For $d_{2}$
the $\zeta_{5}$, present in all but the leading term in the
$N_{f}$-expansion, is now present only in $d_{2}^{(1)}$. In both
cases
the highest $\zeta$-function present cancels and is absent in the
`conformal' $b\rightarrow0$ limit \cite{brodlu}; so $d_{1}^{(0)}$
does
not
involve $\zeta_{3}$ and $d_{2}^{(0)}$ does not involve $\zeta_{5}$.
Since the $\MS$ beta-function coefficients do not
involve $\zeta$-functions, this is presumably not an artefact of the
particular RS chosen but may well be of more fundamental significance.
It may ultimately be connected with the fact that in the
$b\rightarrow 0$ limit the $z$=$2\ell/b$ $UV_{\ell}$ and $IR_{\ell}$
singularities in the Borel plane move off to infinity, leaving only
instanton singularities. As discussed in reference \cite{us} the
$\zeta$-functions are intimately linked with the presence of
renormalon singularities. It is amusing to notice that in the original
NNLO result for $d_{2}$ \cite{origoa3}, which was subsequently found
to be in
error \cite{newrussians1}, $d_{2}^{(0)}$ does contain a non-vanishing
$-\frac{20}{9}\zeta_{5}$ term for SU($3$) QCD. If a fundamental
result about the absence of $\zeta$-functions in the conformal limit
could be established it would have enabled the incorrect result to
have been dismissed at once.

The corresponding results for the $N_{f}$-expansion for the deep
inelastic sum rules are:
\ba
K_{1}&=&-\frac{1}{3}N_{f}+\left(\frac{23}{12}C_{A}-\frac{7}{8}C_{F}
\right)\;,\\
K_{2}&=&N_{f}^{2}\left(\frac{115}{648}\right)+N_{f}\left(-\frac{3535}
{1296}-\frac{\zeta_{3}}{2}+\frac{5}{9}\zeta_{5}\right)C_{A}
+N_{f}\left(\frac{133}{864}+\frac{5}{18}\zeta_{3}\right)C_{F}
\nonumber\\& &+C_{A}^{2}\left(\frac{5437}{648}-
\frac{55}{18}\zeta_{5}\right)
+C_{A}C_{F}\left(-\frac{1241}{432}+\frac{11}{9}\zeta_{3}
\right)+C_{F}^{2}\left(\frac{1}{32}\right)\;.
\ea
In numerical form, for SU($N$) QCD,
\ba
K_{1}&=&-.333N_{f}+\left(1.48N+\frac{.438}{N}\right)\;,\\
K_{2}&=&.177N_{f}^{2}+N_{f}\left(-2.51N-\frac{.244}{N}\right)
+\left(4.53N^{2}+.686+\frac{.008}{N^{2}}\right)\;.
\ea
Expanding in powers of $b$ as in equation (3),
\ba
K_{1}&=&b+\left(\frac{C_{A}}{12}-\frac{7}{8}C_{F}\right)\;,\\
K_{2}&=&b^{2}\left(\frac{115}{72}\right)+b\left(\frac{335}{144}
+\frac{3}{2}\zeta_{3}-\frac{15}{9}\zeta_{5}\right)C_{A}+b\left(
-\frac{133}{288}-\frac{5}{6}\zeta_{3}\right)C_{F}\nonumber\\
& &+C_{A}^{2}\left(-\frac{179}{144}-\frac{11}{4}\zeta_{3}\right)
+C_{A}C_{F}\left(-\frac{389}{192}+\frac{11}{4}\zeta_{3}\right)
+C_{F}^{2}\left(\frac{1}{32}\right)
\;.
\ea
Notice that $K_{1}$ does not contain $\zeta_{3}$; but a similar
remark about the absence of $\zeta_{5}$ in $K_{2}^{(0)}$ holds.

We now wish to demonstrate that the leading term in the $b$-expansion,
when expanded in $N_{f}$, approximates the $N_{f}$-expansion
coefficients well, even in rather low orders.

For $d_{1}$ and $d_{2}$ we have
\ba
d_{1}^{(1)}b&=&.345b=-.115N_{f}+.634N\;,\\
d_{2}^{(2)}b^{2}&=&.776b^{2}=.086N_{f}^{2}-.948N_{f}N+2.61N^{2}\;.
\ea
The subleading, $N$, $N_{f}N$ and $N^{2}$, coefficients approximate
well in sign and magnitude those in the exact expressions in equations
(16) and (17). The leading, $N_{f}$ and $N_{f}^{2}$, coefficients of
course agree exactly.

For $K_{1}$ and $K_{2}$ we have
\ba
K_{1}^{(1)}b&=&b=-.333N_{f}+1.83N\;,\\
K_{2}^{(2)}b^{2}&=&1.59b^{2}=.177N_{f}^{2}-1.95N_{f}N+5.37N^{2}\;.
\ea
The agreement with the exact $N$, $N_{f}N$ and $N^{2}$ coefficients
in equations (22) and (23) is again rather good.

We now turn to a consideration of the RS dependence of the $N_{f}$
and $b$ expansions. In variants of minimal subtraction, where the
$1/\epsilon$ pole in dimensional regularization is subtracted along
with an $N_{f}$-independent finite part, $K$, the QCD perturbative
coefficients will have the form of polynomials in $N_{f}$ as in
equation (2). Modified minimal subtraction ($\MS$),
corresponding to $K$=$(\ln4\pi$--$\gamma_{E})$, with
$\gamma_{E}$=$0.5722\ldots$, Euler's constant, is most commonly
employed. We can consider $\MS$ with
renormalization scale $\mu$=$\mbox{e}^{u}Q$, where $u$ is an
$N_{f}$-independent number. The most general subtraction procedure
which will result in perturbative coefficients polynomial in $N_{f}$,
however, can be regarded as $\MS$ with scale
$\mu$=$\mbox{e}^{u+v/b}Q$, where $v$ is again $N_{f}$-independent.
We shall refer to such renormalization schemes as `regular' schemes.
Of
course the renormalization scheme  is not specified by the scale
and subtraction procedure alone but by higher order beta-function
coefficients as well. Any variant of minimal subtraction with an
$N_{f}$-independent renormalization scale will have $v$=$0$. Momentum
space subtraction (MOM) based on the {\em ggg} vertex at a symmetric
subtraction point $\mu^{2}$=$Q^{2}$ \cite{gbstrs} corresponds to
$u$=$2.56$ and
$v$=$C_{A}f(\xi)$, where $f$ is a cubic polynomial in the gauge
parameter
$\xi$. For the Landau gauge, $\xi$=$0$, $v$=$-2.49C_{A}$. For other
versions of MOM based on the {\em qqg} or ghost vertices, $v$ will
involve $C_{A}$ and $C_{F}$.

Let us denote the perturbative coefficients in the
$\MS$ scheme with $\mu$=$Q$ ($u$=$v$=$0$) by $d_{k}$;
and those with general $u$ and $v$ by
$d^{\prime}_{k}$.
Then
\ba
d^{\prime}_{1}&=&(d_{1}^{(1)}+u)b+(d_{1}^{(0)}+v)\nonumber\\
&=&d_{1}+bu+v\;.
\ea
Changing $v$, one can make the
$d_{1}^{(0)}$ coefficient as large as one pleases and hence destroy
the dominance of the leading-$b$ term noted above for the $D$-function
and the sum rules in low orders; although the leading-$b$ term should
still reproduce asymptotically the $d_{k}^{[k-r]}$ coefficients to
$O(1/k)$ accuracy.

For $d_{2}$ and higher coefficients the specification of the RS will
involve higher beta-function coefficients as well as the scale and
subtraction procedure. The RG-improved coupling, $a(\mu^{2})$, will
evolve with renormalization scale according to the beta-function
equation
\be
\frac{\mbox{d}a}{\mbox{d}\ln\mu}=-ba^{2}(1+ca+c_{2}a^{2}+\cdots+
c_{k}a^{k}+\cdots)
\;.
\ee
Here $b$ and $c$ are universal with \cite{vold}
\ba
b&=&\frac{1}{6}(11C_{A}-2N_{f})\;,\nonumber\\
c&=&\left[-\frac{7}{8}\frac{C_{A}^{2}}{b}-
\frac{11}{8}\frac{C_{A}C_{F}}{b}+\frac{5}{4}C_{A}
+\frac{3}{4}C_{F}\right]\;.
\ea
Integrating equation (31) with a suitable choice of boundary
condition \cite{spew}, one obtains a transcendental equation for $a$:
\be
b\ln\frac{\mu}{\Lambda}=\frac{1}{a}+c\ln\frac{ca}{1+ca}+
\int_{0}^{a}\mbox{d}x\left[-\frac{1}{x^{2}B(x)}+\frac{1}{x^{2}(1+cx)}
\right]\;,
\ee
where $B(x)$=$(1+cx+c_{2}x^{2}+c_{3}x^{3}+\cdots+c_{k}x^{k}+\cdots)$.
The beta-function coefficients, $c_{2}$, $c_{3}$, $\ldots$ together
with $b\ln\frac{\mu}{\Lambda}$ label the RS. In a fixed order
perturbative calculation one would truncate the beta-function. For the
all-orders resummations of the next section, however, one requires
an all-orders definition of the coupling. In the
$\MS$ scheme the higher beta-function coefficients,
$c_{2}^{\MS}$,
$c_{3}^{\MS}$,
$\ldots$,
presumably exhibit factorial growth,
$c_{k}^{\MS}\sim k!$;
and the `$a$' coupling in the
Borel integral would not
be defined, since $B(x)$ would itself need to be defined by a Borel
integral or other summation. One therefore needs to use a finite
scheme \cite{thooft} where $B(x)$ has a finite radius of convergence
and can be
summed. An extreme example is the so-called 't Hooft scheme
\cite{thooft} where
$c_{2}$=$c_{3}$=$\cdots$=$c_{k}$=$\cdots$=$0$, $B(x)$=$1+cx$. This
results in the all-orders definition of the coupling,
\be
b\ln\frac{\mu}{\Lambda}=\frac{1}{a}+c\ln\frac{ca}{1+ca}\;.
\ee
In such a finite scheme, where $c_{2}$, $c_{3}$, $\ldots$ are
$N_{f}$-independent, the $b$-expansion of equation~(3) will contain an
extra $d_{k}^{(-1)}/b$ term (for $k>1$) and the $d_{k}$ will strictly
no longer be polynomials in $N_{f}$. $bd_{k}$, however, is a
polynomial
in $N_{f}$ of degree $k+1$. The leading-$b$ coefficients,
$d_{k}^{(k)}$,
in regular schemes are independent of $c_{2}$, $c_{3}$, $\ldots$,
since these
beta-function coefficients, $c_{k}$, are $O(1/N_{f})$ relative to
$d_{k}$.
\section{Leading-$b$ Resummations}

Let us begin by recalling the definition of the Borel transform. If
$D$ has a series expansion in `$a$' as in equation (1), we can write
\be
D=\int_{0}^{\infty}\mbox{d}z\,\mbox{e}^{-z/a}B[D](z)
\;,
\ee
where $B[D](z)$ is the Borel transform of $D$, defined by ($d_{0}$=1)
\be
B[D](z)=\sum_{m=0}^{\infty}\frac{z^{m}d_{m}}{m!}\;.
\ee
For the Euclidean quantities, $\tilde{D}$ and $\tilde{K}$, defined
earlier we shall deduce from the exact large-$N_{f}$ results that,
in the $\MS$ scheme with $\mu$=$\mbox{e}^{-5/6}Q$,\,
$B[D](z)$ is of the form
\ba
B[D](z)&=&\sum_{\ell=1}^{\infty}\frac{A_{0}(\ell)+A_{1}(\ell)z+
\overline{A}_{1}(\ell)z+\overline{A}_{2}(\ell)z^{2}+\cdots}{\left(1+
\frac{z}{z_{\ell}}\right)^{\alpha_{\ell}+\overline{\alpha}_{\ell}}}
\nonumber\\
& &+\sum_{\ell=1}^{\infty}\frac{B_{0}(\ell)+B_{1}(\ell)z+
\overline{B}_{1}(\ell)z+\overline{B}_{2}(\ell)z^{2}+\cdots}{\left(1-
\frac{z}{z_{\ell}}\right)^
{\beta_{\ell}+\overline{\beta}_{\ell}}}\;+\cdots\;,
\ea
where $z_{\ell}$=$2\ell/b$. The two terms correspond to a
summation over the ultra-violet renormalons, $UV_{\ell}$, and
infra-red
renormalons, $IR_{\ell}$, respectively. $A_{0}(\ell)$, $A_{1}(\ell)$,
$\alpha_{\ell}$ and $B_{0}(\ell)$, $B_{1}(\ell)$, $\beta_{\ell}$ will
be obtained from the large-$N_{f}$ results. The barred terms are
sub-leading in $N_{f}$ and remain unknown. The use of the so-called
`V-scheme' \cite{blm}, $\MS$ with
$\mu$=$\mbox{e}^{-5/6}Q$, means that only the constant and $O(z)$
terms
in the numerator polynomials are leading in $N_{f}$. For a general
$\MS$ scale, $\mu$=$\mbox{e}^{u}Q$, an overall factor
$\mbox{e}^{bz(u+5/6)}$ should multiply the unbarred leading-$N_{f}$
terms in the numerator. The presence of this exponential factor, when
it is expanded in powers of $z$, can mask the presence of the $UV$
and $IR$ renormalons in low orders of perturbation theory.

Notice that, whilst the residue at each renormalon singularity is
only
known to leading order in $N_{f}$, the $A_{0}(\ell)$ and
$B_{0}(\ell)$ constant terms in the numerator polynomials are known
exactly; indeed
$\sum_{\ell=1}^{\infty}(A_{0}(\ell)+B_{0}(\ell))$=$1$, as
is required to reproduce the unit coefficient of the $O(a)$ term in
$D$ in equation (1). We now turn to the explicit determination of
the coefficients and exponents for $\tilde{D}$ and $\tilde{K}$.

For $\tilde{D}$ the leading-$N_{f}$ terms in the QED Gell-Mann--Low
function are generated by \cite{broad}
\[
\psi_{n}^{[n]}=\frac{3^{2-n}}{2}\left(\frac{\mbox{d}}
{\mbox{d}x}\right)^{n-2}P(x)\biggr|_{x=1}\;,
\]
where
\be
P(x)=\frac{32}{3(1+x)}\sum_{k=2}^{\infty}\frac{(-1)^{k}k}{(k^{2}
-x^{2})^{2}}\;.
\ee
In the V-scheme, $\MS$ with
$\mu$=$\mbox{e}^{-5/6}Q$, one then has
\be
d_{n}^{(n)}=\left(-\frac{3}{2}\right)^{n}2\psi_{n+2}^{[n+2]}\;.
\ee
It is then straightforward to deduce that the coefficients and
exponents in equation (37) for $B[\tilde{D}](z)$ are
\begin{eqnarray*}
A_{0}(\ell)&=&\frac{8}{3}\frac{(-1)^{\ell+1}(3\ell^{2}+6\ell+2)}
{\ell^{2}(\ell+1)^{2}(\ell+2)^{2}},\;\;
A_{1}(\ell)=\frac{8}{3}\frac{b(-1)^{\ell+1}(\ell+\frac{3}{2})}
{\ell^{2}(\ell+1)^{2}(\ell+2)^{2}}\nonumber\\
\ell&=&1,2,3,\ldots
\end{eqnarray*}
\[
\begin{array}{llll}
B_{0}(1)=0,&B_{0}(2)=1,&B_{0}(\ell)=-A_{0}(-\ell)&\ell\geq3\\
B_{1}(1)=0,&B_{1}(2)=0,&B_{1}(\ell)=-A_{1}(-\ell)&\ell\geq3\\
\end{array}
\]
\be
\alpha_{\ell}=2\;\;\ell=1,2,3,\ldots,\;\;\;\;
\beta_{2}=1,\;\beta_{\ell}=2
\;\;\;\ell\geq3\;.
\ee
So $IR_{1}$ is absent, as required from the absence of a dimension
two condensate in the OPE \cite{beneke,us}. $IR_{2}$ is a single
pole.
All the
other singularities are double poles. Not only are the coefficients
for the $UV_{\ell}$ and $IR_{\ell}$ singularities related by the
curious symmetry $B_{0,1}(\ell)$=$-A_{0,1}(-\ell)$; but the form of
$A_{0}(\ell)$ means that there is an additional relation,
$A_{0}(\ell)$=$-B_{0}(\ell+2)$, so that the constant term in the
numerator polynomial for $UV_{\ell}$ exactly cancels that for
$IR_{\ell+2}$. This ensures that
\[
\sum_{\ell=1}^{\infty}(A_{0}(\ell)+B_{0}(\ell))=B_{0}(2)=1
\;,
\]
which, as noted earlier, is required to reproduce the unit
coefficient of the $O(a)$ term in the perturbative expansion. The
precise origin of these relations between $UV$ and $IR$ renormalons
remains unclear and deserves further study. They have also been noted
and discussed in reference \cite{benthes}.

For the $D$-function the singularity nearest the origin is $UV_{1}$
and from the $A_{0}(1)$, $A_{1}(1)$ in equation (40) this should
correspond to
\be
d_{n}^{(n)}|_{UV_{1}}=\frac{12n+22}{27}n!\left(-\frac{1}{2}\right)^{n}
\;.
\ee
\begin{table}
\begin{center}
\begin{tabular}{|l|l|l|l|l|}\hline
\multicolumn{1}{|c|}{$n$}
&\multicolumn{1}{c|}{$d_{n}^{(n)}$}
&\multicolumn{1}{c|}{$UV_{1}$}
&\multicolumn{1}{c|}{$UV$}
&\multicolumn{1}{c|}{$IR$}\\\hline
\zrow{ 0}{\mbox{ }1       }{\mbox{ }.8148148    }
{\mbox{ }.7198242    }{\mbox{ }.28018     }
\zrow{ 1}{-.4874471       }{-.6296296           }
{-.5921448           }{\mbox{ }.10470     }
\zrow{ 2}{\mbox{ }.8938293}{\mbox{ }.8518519    }
{\mbox{ }.8258924    }{\mbox{ }.06794     }
\zrow{ 3}{-1.525257       }{-1.611111           }
{-1.586113           }{\mbox{ }.06086     }
\zrow{ 4}{\mbox{ }3.927235}{\mbox{ }3.888889    }
{\mbox{ }3.858335    }{\mbox{ }.06890     }
\zrow{ 5}{-11.24973       }{-11.38889           }
{-11.34378           }{\mbox{ }.09405     }
\zrow{ 6}{\mbox{ }39.23893}{\mbox{ }39.16667    }
{\mbox{ }39.08871    }{\mbox{ }.15022     }
\zrow{ 7}{-154.1541       }{-154.5833           }
{-154.4291           }{\mbox{ }.27499     }
\zrow{ 8}{\mbox{ }688.5574}{\mbox{ }688.3333    }
{\mbox{ }687.9894    }{\mbox{ }.56801     }
\zrow{ 9}{-3410.339       }{-3412.500           }
{-3411.647           }{\mbox{ }1.3079     }
\zrow{10}{\mbox{ }18638.50}{\mbox{ }18637.50    }
{\mbox{ }18635.17    }{\mbox{ }3.3243     }
\hline
\end{tabular}
\caption{Leading-$b$ coefficients, $d_{n}^{(n)}$, for the Adler
$D$-function, $\tilde{D}$, compared with the contribution of the
first
$UV$ renormalon, `$UV_{1}$', (equation (41)). The V-scheme,
$\MS$ with $\mu$=$\mbox{e}^{-5/6}Q$, is assumed.
`$UV$' and `$IR$' denote the separate sums over the $UV_{\ell}$ and
$IR_{\ell}$ singularities.}
\end{center}
\end{table}

In Table 1 we compare the exact leading-$b$, $d_{n}^{(n)}$,
coefficients with the contribution from $UV_{1}$ of equation (41).
$UV$ and $IR$ denote the separate sums over the $UV_{\ell}$ and
$IR_{\ell}$ singularities. The $\MS$ scheme with
$\mu$=$\mbox{e}^{-5/6}Q$ (V-scheme) is assumed. With this choice of
scheme $UV_{1}$ dominates even in low orders and the alternating
factorial behaviour is apparent.

We now consider the coefficients and exponents in equation (37) for
$B[\tilde{K}](z)$. The generating function for the leading-$b$
coefficient in the V-scheme is \cite{broadkat}
\be
K_{n}^{(n)}=\frac{1}{3}\left(\frac{1}{2}\right)^{n}
\frac{\mbox{d}^{n}}{\mbox{d}x^{n}}\frac{(3+x)}{(1-x^{2})(1-
\frac{x^{2}}{4})}\biggr|_{x=0}
\;.
\ee
This results in
\be
B[\tilde{K}](z)=\frac{\frac{4}{9}}{(1+\frac{bz}{2})}-
\frac{\frac{1}{18}}{(1+\frac{bz}{4})}+
\frac{\frac{8}{9}}{(1-\frac{bz}{2})}-
\frac{\frac{5}{18}}{(1-\frac{bz}{4})}
\;.
\ee

The terms correspond to $UV_{1}$,  $UV_{2}$, $IR_{1}$, $IR_{2}$
respectively. Each numerator and exponent will contain in addition
$O(1/N_{f})$ corrections corresponding to the barred terms in equation
(37). The constant terms in the numerators sum to 1, again
ensuring a unit $O(a)$ coefficient in $\tilde{K}$. It would be
interesting to try to understand the fact that only the first two
$UV$ and $IR$ renormalons are leading in $N_{f}$ in the context of
the OPE for the deep inelastic sum rules, a topic discussed in
reference \cite{gg}.

$K_{n}^{(n)}$ is then given by (in the V-scheme)
\be
K_{n}^{(n)}=\frac{8}{9}n!\left(\frac{1}{2}\right)^{n}+
\frac{4}{9}n!\left(-\frac{1}{2}\right)^{n}-
\frac{5}{18}n!\left(\frac{1}{4}\right)^{n}-
\frac{1}{18}n!\left(-\frac{1}{4}\right)^{n}\;.
\ee
\begin{table}
\begin{center}
\begin{tabular}{|l|l|l|l|l|}\hline
\multicolumn{1}{|c|}{$n$}
&\multicolumn{1}{c|}{$K_{n}^{(n)}$}
&\multicolumn{1}{c|}{$UV_{1}+IR_{1}$}
&\multicolumn{1}{c|}{$UV$}
&\multicolumn{1}{c|}{$IR$}\\\hline
\zrow{ 0}{1              }{1.333333       }
{\mbox{ }.3888889      }{.6111111       }
\zrow{ 1}{.1666667       }{.2222222       }
{-.2083333      }{.3750000       }
\zrow{ 2}{.6250000       }{.6666667       }
{\mbox{ }.2152778      }{.4097222       }
\zrow{ 3}{.3125000       }{.3333333       }
{-.3281250      }{.6406250       }
\zrow{ 4}{1.968750       }{2.000000       }
{\mbox{ }.6614583      }{1.307292       }
\zrow{ 5}{1.640625       }{1.666667       }
{-1.660156      }{3.300781       }
\zrow{ 6}{14.94141       }{15.00000       }
{\mbox{ }4.990234      }{9.951172       }
\zrow{ 7}{17.43164       }{17.50000       }
{-17.48291      }{34.91455       }
\zrow{ 8}{209.7949       }{210.0000       }
{\mbox{ }69.96582      }{139.8291       }
\zrow{ 9}{314.6924       }{315.0000       }
{-314.9231      }{629.6155       }
\zrow{10}{4723.846       }{4725.000       }
{\mbox{ }1574.808      }{3149.039       }
\hline
\end{tabular}
\caption{As for Table 1 but for the Deep Inelastic Scattering sum
rules,
$\tilde{K}$, $K_{n}^{(n)}$. `$UV_{1}$+$IR_{1}$' denotes the
contribution of the singularities nearest the origin, the first two
terms of equation~(44).}
\end{center}
\end{table}
Table 2 shows that $K_{n}^{(n)}$ is dominated, even in low orders, by
the combined $UV_{1}+IR_{1}$ contributions of the two singularities
nearest the origin, the first two terms of equation (44).

We now wish to use the Borel integrals to perform the leading-$b$
resummation defined in equations (5,6). For the Adler $D$-function we
have
\ba
\tilde{D}^{(L)}(a)&=&\int_{0}^{\infty}\mbox{d}z\,\mbox{e}^{-F(a)z}
\sum_{\ell=1}^{\infty}\frac{A_{0}(\ell)+A_{1}(\ell)z}{(1+
\frac{z}{z_{\ell}})^{2}}\nonumber\\
& &+\int_{0}^{\infty}\mbox{d}z\,\mbox{e}^{-F(a)z}\left(\frac{B_{0}(2)}
{(1-\frac{z}{z_{2}})}+
\sum_{\ell=3}^{\infty}\frac{B_{0}(\ell)+B_{1}(\ell)z}{(1-
\frac{z}{z_{\ell}})^{2}}\right)
\;.
\ea
The coefficients $A_{0}$, $A_{1}$, $B_{0}$, $ B_{1}$ are summarised
in
equation (40). We assume that the resummation has been performed in a
finite RS corresponding to $\MS$ subtraction with
$\mu$=$\mbox{e}^{u+v/b}Q$. `$a$', the coupling, is then defined by
the
integrated beta-function equation of equation (33). One then finds
that
the exponent in equation (45) is
\be
F(a)=b\ln\frac{Q}{\Lambda_{\MS}}-\frac{5}{6}b-
c\ln\frac{ca}{1+ca}
+v-\int_{0}^{a}\mbox{d}x\left[-\frac{1}{x^{2}B(x)}+
\frac{1}{x^{2}(1+cx)}\right]\;.
\ee
The RS dependence of $\tilde{D}^{(L)}$ reflects the fact that only a
subset of the perturbation series has been resummed, hence violating
the exact RS-invariance which would apply to the full series. We shall
return to this RS dependence in a moment.

The first, $UV$ renormalon, term in equation (45) is a completely
well-defined integral. It may be performed in terms of the exponential
integral function (with negative argument),
\be
\mbox{Ei}(x)=-\int_{-x}^{\infty}\mbox{d}t\frac{\mbox{e}^{-t}}{t}
\;.
\ee
The first term yields
\ba
\tilde{D}^{(L)}(a)|_{UV}&=&\sum_{\ell=1}^{\infty}
z_{\ell}\{\mbox{e}^{F(a)z_{\ell}}
\mbox{Ei}(-F(a)z_{\ell})\left[F(a)z_{\ell}(A_{0}(\ell)-z_{\ell}
A_{1}(\ell))-z_{\ell}A_{1}(\ell)\right]\nonumber\\
& &+(A_{0}(\ell)-z_{\ell}A_{1}(\ell))\}\;.
\ea
To evaluate the second, $IR$ renormalon, term we shall use a principal
value prescription; correspondingly we need to define $\mbox{Ei}(x)$
with a positive argument as a principal value. We find
\ba
\tilde{D}^{(L)}(a)|_{IR}&=&\mbox{e}^{-F(a)z_{2}}z_{2}B_{0}(2)\mbox{Ei}
(F(a)z_{2})\nonumber\\
& &+
\sum_{\ell=3}^{\infty}z_{\ell}\{\mbox{e}^{-F(a)z_{\ell}}
\mbox{Ei}(F(a)z_{\ell})\left[F(a)z_{\ell}(B_{0}(\ell)+z_{\ell}
B_{1}(\ell))-z_{\ell}B_{1}(\ell)\right]\nonumber\\
& &-(B_{0}(\ell)+z_{\ell}B_{1}(\ell))\}\;.
\ea
Finally
\be
\tilde{D}^{(L)}(a)=\tilde{D}^{(L)}(a)|_{UV}+
\tilde{D}^{(L)}(a)|_{IR}\;.
\ee

For the Deep Inelastic Scattering sum rules we will have,
analogously,
using equation (43),
\ba
\tilde{K}^{(L)}(a)&=&\int_{0}^{\infty}\mbox{d}z\,\mbox{e}^{-F(a)z}
\left(\frac{\frac{4}{9}}{(1+\frac{z}{z_{1}})}-
\frac{\frac{1}{18}}{(1+\frac{z}{z_{2}})}\right)\nonumber\\& &+
\int_{0}^{\infty}\mbox{d}z\,\mbox{e}^{-F(a)z}
\left(\frac{\frac{8}{9}}{(1-\frac{z}{z_{1}})}-
\frac{\frac{5}{18}}{(1-\frac{z}{z_{2}})}\right)\;.
\ea
These integrals may be expressed once again in terms of
$\mbox{Ei}(x)$:
\be
\tilde{K}^{(L)}(a)|_{UV}=\left[-\frac{4}{9}\mbox{e}^{F(a)z_{1}}
z_{1}\mbox{Ei}(-F(a)z_{1})+\frac{1}{18}\mbox{e}^{F(a)z_{2}}
z_{2}\mbox{Ei}(-F(a)z_{2})\right]
\ee
and
\be
\tilde{K}^{(L)}(a)|_{IR}=\left[\frac{8}{9}\mbox{e}^{-F(a)z_{1}}
z_{1}\mbox{Ei}(F(a)z_{1})-\frac{5}{18}\mbox{e}^{-F(a)z_{2}}
z_{2}\mbox{Ei}(F(a)z_{2})\right]\;.
\ee
Similarly
\be
\tilde{K}^{(L)}(a)=\tilde{K}^{(L)}(a)|_{UV}+
\tilde{K}^{(L)}(a)|_{IR}\;.
\ee
Before we numerically evaluate these results and comment further on
RS dependence, we shall derive the analogous resummations for the
Minkowski continuations of the $D$-function, the $e^{+}e^{-}$
annihilation $R$-ratio and the analogous quantity in $\tau$-decay,
$R_{\tau}$. The $R$-ratio is related to $D$ by a dispersion
relation,
\be
R(s)=\frac{1}{2\pi i}\int_{-s-i\epsilon}^{-s+i\epsilon}
\mbox{d}Q^{2}\frac{D(Q^{2})}{Q^{2}}\;.
\ee
Here $s$ is the physical timelike Minkowski squared momentum
transfer. A perturbative result for $R$ of the form of equation (9)
can be written down involving a quantity $\tilde{R}$ with
perturbative coefficients $r_{k}$. The $r_{k}$ are directly related
to the $d_{k}$ via the dispersion relation (55): $r_{1}$=$d_{1}$,
$r_{2}$=$d_{2}-\pi^{2}b^{2}/12$. The $\pi^{2}$ term arises due to
analytical continuation. In the Borel plane one finds (to leading
order in $N_{f}$)
\be
B[\tilde{R}](z)=\frac{\sin(\pi bz/2)}{\pi bz/2}B[\tilde{D}](z)\;.
\ee
The leading-$b$ resummation is then obtained from equation (45)
simply by adding an extra $\frac{\sin(\pi bz/2)}{\pi bz/2}$ factor in
the integrand.
\ba
\tilde{R}^{(L)}(a)&=&\int_{0}^{\infty}\mbox{d}z\,\mbox{e}^{-F(a)z}
\,\frac{\sin(\pi bz/2)}{\pi bz/2}
\sum_{\ell=1}^{\infty}\frac{A_{0}(\ell)+A_{1}(\ell)z}{(1+
\frac{z}{z_{\ell}})^{2}}\nonumber\\
& &+\int_{0}^{\infty}\mbox{d}z\,\mbox{e}^{-F(a)z}
\,\frac{\sin(\pi bz/2)}{\pi bz/2}
\left(\frac{B_{0}(2)}
{(1-\frac{z}{z_{2}})}+
\sum_{\ell=3}^{\infty}\frac{B_{0}(\ell)+B_{1}(\ell)z}{(1-
\frac{z}{z_{\ell}})^{2}}\right)\;.
\ea
Writing the `sin' as a sum of complex exponentials and using partial
fractions, the $UV$ integrals can be explicitly performed in terms
of the generalised exponential integral functions $\mbox{Ei}(n,w)$,
with complex argument $w$, defined for $\mbox{Re}\,w>0$ by
\be
\mbox{Ei}(n,w)=\int_{1}^{\infty}\mbox{d}t\frac{\mbox{e}^{-wt}}{t^{n}}
\;.
\ee
One also needs
\be
\int_{0}^{\infty}\mbox{d}z\,\mbox{e}^{-F(a)z}\,\frac{\sin(\pi bz/2)}
{z}=\arctan\left(\frac{\pi b}{2F(a)}\right)\;.
\ee
One finds
\ba
\tilde{R}^{(L)}(a)|_{UV}&=&
\frac{2}{\pi b}\left(\frac{8\zeta_{2}}{3}-\frac{11}{3}\right)
\arctan\left(\frac{\pi b}{2F(a)}\right)
\nonumber\\
& &+\frac{2}{\pi b}\sum_{\ell=1}^{\infty}\biggr\{
A_{0}(\ell)\phi_{+}(1,\ell)
+(A_{0}(\ell)-A_{1}(\ell)z_{\ell})\phi_{+}(2,\ell)\biggr\}\;,
\ea
where
\be
\phi_{+}(p,q)=\mbox{e}^{F(a)z_{q}}(-1)^{q}
{\rm Im}[{\rm Ei}(p,F_{+}z_{q})]
\ee
with $F_{\pm}$=$F(a)\pm\frac{i\pi b}{2}$.

To evaluate the principal value of the $IR$ contribution in
equation~(57) one needs to continue Ei($n,w$), defined by
equation~(58) for $\mbox{Re}\,w>0$, to $\mbox{Re}\,w<0$. With the
standard continuation one then arrives at a function analytic
everywhere in the cut complex $w$-plane, except at $w$=0; and with a
branch cut running along the negative real axis. Explicitly
\cite{abst}
\be
{\rm Ei}(n,w)=\frac{(-w)^{n-1}}{(n-1)!}\left[-\ln w-\gamma_{E}+
\sum_{m=1}^{n-1}\frac{1}{m}\right]-\sum_{
\stackrel{\scriptstyle m=0}
{m\neq n-1}}^{\infty}\frac{(-w)^{m}}
{(m-n+1)m!}\;,
\ee
with $\gamma_{E}$=$0.5722\ldots$, Euler's constant. The $\ln w$ term
in equation~(62) means that Ei($n,w$) is not a real function. For
instance, for negative real $w$ one has
Ei$(1,-x\pm i\epsilon)$=Ei$(x)\mp i\pi$, where Ei$(x)$ is the
principal value of equation~(47) used to define the $IR$ renormalon
contribution for the Euclidean quantities.

In order to evaluate the $IR$ renormalon contribution correctly,
one in fact
needs to continue Ei($n,w$) as a real function, so that for
$\mbox{Re}\,w<0$ one makes the replacement
$\ln w\rightarrow\ln w+i\pi\,{\rm sign}({\rm Im}\,w)$.
Correspondingly,
one should define the $IR$ analogue of equation~(61),
\be
\phi_{-}(p,q)=\mbox{e}^{-F(a)z_{q}}(-1)^{q}
{\rm Im}[{\rm Ei}(p,-F_{+}z_{q})]
-\frac{\mbox{e}^{-F(a)z_{q}}(-1)^{q}z_{q}^{p-1}}{(p-1)!}\pi{\rm Re}
[(F_{+})^{p-1}]\;,
\ee
where Ei$(p,-F_{+}z_{q})$ is defined by equation~(62). The principal
value of the $IR$ renormalon contribution is then given by
\ba
\tilde{R}^{(L)}(a)|_{IR}&=&
\frac{2}{\pi b}\left(\frac{14}{3}-\frac{8\zeta_{2}}{3}\right)
\arctan\left(\frac{\pi b}{2F(a)}\right)
+\frac{2B_{0}(2)}{\pi b}\phi_{-}(1,2)\nonumber\\& &
+\frac{2}{\pi b}\sum_{\ell=3}^{\infty}\biggr\{
B_{0}(\ell)\phi_{-}(1,\ell)
+(B_{0}(\ell)+B_{1}(\ell)z_{\ell})\phi_{-}(2,\ell)\biggr\}.
\ea
Then
\be
\tilde{R}^{(L)}(a)=\tilde{R}^{(L)}(a)|_{UV}+
\tilde{R}^{(L)}(a)|_{IR}\;.
\ee

The $\tau$-decay analogue of the $R$-ratio, $R_{\tau}$, can be
defined
in terms of the $R$-ratio by the integral representation \cite{bnp}
\ba
R_{\tau}&=&2\int_{0}^{M_{\tau}^{2}}\frac{\mbox{d}s}{M_{\tau}^{2}}
(1-s/M_{\tau}^{2})^{2}(1+2s/M_{\tau}^{2})\hat{R}(s)\nonumber\\
&=&d(R)(|V_{ud}|^{2}+|V_{us}|^{2})\left[1+
\frac{3}{4}C_{F}\tilde{R}_{\tau}\right]
\;.
\ea
Here $\hat{R}(s)$ denotes $R(s)$ with the $\sum_{f}Q_{f}^{2}$
replaced
by $|V_{ud}|^{2}+|V_{us}|^{2}\approx1$, where the $V$'s are KM mixing
matrix elements. $\tilde{R}_{\tau}$ has the form
\be
\tilde{R}_{\tau}=a+r_{1}^{\tau}a^{2}+r_{2}^{\tau}a^{3}+\cdots+
r_{k}^{\tau}a^{k+1}+\cdots
\ee
It is then straightforward to show that
\be
B[\tilde{R}_{\tau}](z)=\frac{\sin(\pi bz/2)}{\pi bz/2}\left[
\frac{2}{(1-\frac{bz}{2})}-\frac{2}{(1-\frac{bz}{6})}+
\frac{1}{(1-\frac{bz}{8})}\right]B[\tilde{D}](z)
\;.
\ee
Proceeding in a manner analogous to that for the $R$-ratio,
we find
\ba
\tilde{R}_{\tau}^{(L)}(a)|_{UV}&=&
\frac{2}{\pi b}\left(\frac{8\zeta_{2}}{3}-\frac{11}{3}\right)
\arctan\left(\frac{\pi b}{2F(a)}\right)\nonumber\\
& &+\frac{4}{\pi b}\sum_{\ell=1}^{\infty}
[(A_{0}(\ell)(G(\ell)+
H(\ell))-z_{\ell}A_{1}(\ell)G(\ell))\phi_{+}(1,\ell)
\nonumber\\& &+H(\ell)(A_{0}(\ell)-z_{\ell}A_{1}(\ell))
\phi_{+}(2,\ell)]
\ea
and
\ba
\tilde{R}_{\tau}^{(L)}(a)|_{IR}&=&
\frac{2}{\pi b}\left(\frac{14}{3}-\frac{8\zeta_{2}}{3}\right)
\arctan\left(\frac{\pi b}{2F(a)}\right)\nonumber\\
& &+\frac{4}{\pi b}\left(-\frac{14}{3}+\frac{64}{3}\ln 2-
8\zeta_{3}+bz_{1}\left(\frac
{23}{3}-\frac{32}{3}\ln 2\right)\right)\phi_{-}(1,1)\nonumber\\
& &-\frac{12}{\pi b}\phi_{-}(1,2)+\frac{4}{\pi b}
\left(-\frac{703}{18}+64\ln 2+bz_{3}\left(\frac{245}{36}-
\frac{32}{3}\ln 2\right)\right)\phi_{-}(1,3)\nonumber\\
& &+\frac{4}{\pi b}\left(-\frac{1627}{972}-\frac{128}{81}
\ln 2+bz_{4}\left(\frac{2035}{7776}+
\frac{16}{81}\ln 2\right)\right)\phi_{-}(1,4)\nonumber\\
& &+\frac{4}{\pi b}\left(-\frac{11}{27}+
\frac{bz_{3}}{6}\right)\phi_{-}(2,3)+\frac{4}{\pi b}
\left(\frac{247}{648}-\frac{5bz_{4}}{162}\right)\phi_{-}(2,4)
\nonumber\\
& &-\frac{8}{\pi b}\left(-\frac{11}{27}+\frac{bz_{3}}{18}
\right)\phi_{-}(3,3)
+\frac{8}{\pi b}\left(\frac{13}{432}-\frac{5bz_{4}}{1728}
\right)\phi_{-}(3,4)\nonumber\\
& &+\frac{4}{\pi b}\sum_{\ell=5}^{\infty}[(B_{0}(\ell)
(G(-\ell)+H(-\ell))+z_{\ell}B_{1}(\ell)G(-\ell))
\phi_{-}(1,\ell)
\nonumber\\& &+H(-\ell)(B_{0}(\ell)+z_{\ell}B_{1}(\ell))
\phi_{-}(2,\ell)]
\;,
\ea
where
\[
G(\ell)=\frac{6\ell(3\ell^{2}+16\ell+19)}{(\ell+1)^{2}
(\ell+3)^{2}(\ell+4)^{2}},\;\;\;
H(\ell)=\frac{6}{(\ell+1)(\ell+3)(\ell+4)}\;.
\]
Then, as before,
\be
\tilde{R}_{\tau}^{(L)}(a)=\tilde{R}_{\tau}^{(L)}(a)|_{UV}+
\tilde{R}_{\tau}^{(L)}(a)|_{IR}\;.
\ee

Before we proceed to discuss RS dependence further and to evaluate
numerically these resummed expressions, we would like to make some
remarks. The first concerns the ease of evaluation of both the
Euclidean resummed expressions, equations~(48), (49) and (52), (53),
and those for the Minkowski quantities, equations (60), (64) and
(69),
(70). Even though these expressions contain infinite summations over
the contributions for the $UV_{\ell}$ and $IR_{\ell}$ singularities,
successive terms in the sums are strongly damped, with the result
that, in order to obtain the three significant figure accuracy of the
resummed results to be tabulated in the next section, it is only
necessary to retain terms up to and including $\ell$=$7$ in each sum.
The resummations can then be straightforwardly and rapidly evaluated.

The second remark concerns the connection between these explicit
expressions for the principal value of the Borel sum and the
inequivalent continuations of the running coupling representation to
the Minkowski region in reference \cite{neubert}. It is
straightforward to show that procedure `1' of reference
\cite{neubert} for $\tilde{R}$ and $\tilde{R_{\tau}}$ corresponds
exactly to evaluating equations~(60), (64) and (69), (70) using
$\phi_{-}(p,q)$ defined by equation~(63) with the second term
omitted, i.e. using the standard continuation of Ei($n,w$) defined
in equation~(62). This does not produce the principal value of the
Borel sum. Worse still, the Borel sum contains pieces involving
single $IR$ renormalon poles together with a $\sin\frac{\pi bz}{2}$
factor, which are well-defined and finite due to the compensating
zero contained in the `sin'. These contributions, which do not
require regulation, are evaluated incorrectly with the standard
continuation. Procedure `2' of reference \cite{neubert} corresponds
to evaluating the Borel sum, incorrectly omitting the second term
in equation~(63) for some of the $IR$ singularities and, correctly,
retaining it for others. In our view the inequivalent continuations
of the running coupling representation of reference \cite{neubert} to
the Minkowski region correspond to various ways of wrongly evaluating
the Borel sum. We see no reason to believe that these discrepancies
have a physical relevance, or that they reflect inadequacies in the
definition of the OPE in the Minkowski region, as suggested in
reference \cite{neubert}. We agree with reference \cite{ballbenbraun}
that, with our present state of knowledge, the regulated Borel sum
provides a satisfactory framework for combining $IR$ renormalon
ambiguities with the vacuum condensate ambiguities in the OPE.

\section{RS-dependence of the resummed results}

Armed with these resummed expressions, we now return to the question
of the RS dependence of $D^{(L)}(a)$ before presenting the numerical
results.

For a generic quantity we can write
\be
D=D^{(L)}(a)+D^{(NL)}(a)\;.
\ee
`$D$' on the left of the equation denotes the full Borel sum, i.e.
equation (37) with barred and unbarred terms included and the
$IR$ singularities principal value regulated; and we assume that this
exists. Crucially, the full Borel sum is RS independent and so will
not depend on `$a$'. The L and NL components do depend on `$a$',
however. We can consider `$a$' varying between $a$=$0$ and
$a$=$+\infty$,
labelling possible RS's. From equation (46), as $a\rightarrow0$ so
$F(a)\rightarrow+\infty$, resulting from the $-c\ln\frac{ca}{1+ca}$
term, and we have assumed $c>0$, which is true for $N_{f}\leq8$ in
SU($3$) QCD. One then has $D^{(L)}(0)$=$0$.
Correspondingly, from equation (72), $D^{(NL)}(0)$=$D$ and so, as
$a\rightarrow0$, the NL component contributes the whole resummed $D$.
As `$a$' increases the $-c\ln\frac{ca}{1+ca}$ term in equation (46)
decreases and as $a\rightarrow\infty$ it vanishes, resulting in a
finite limit $F(\infty)$:
\be
F(\infty)=b\ln\frac{Q}{\Lambda_{\MS}}-\frac{5}{6}b+v-
\int_{0}^{\infty}\mbox{d}x\left[-\frac{1}{x^{2}B(x)}+
\frac{1}{x^{2}(1+cx)}\right]\;.
\ee
We assume that $B(x)$ is such that the integral exists. Thus
$D^{(L)}(a)$ increases from $D^{(L)}(0)$=$0$ to a finite maximum
value,
$D^{(L)}(\infty)$, as `$a$' increases; correspondingly, $D^{(NL)}(a)$,
which provides the whole resummed $D$ at $a$=$0$, decreases as `$a$'
increases.

This RS dependence of $D^{(L)}(a)$ is clearly problematic. It is
monotonic and hence there is no basis for choosing a particular
scheme. There is also a dependence on the particular finite scheme,
characterised by the choice of $B(x)$, and on the parameter $v$. The
maximum value, $D^{(L)}(\infty)$, does perhaps minimize the relative
contribution of the unknown $D^{(NL)}$ component but there is no
guarantee that  $D^{(NL)}(\infty)$ is positive; and it is entirely
possible that $D^{(L)}(\infty)$ overestimates $D$.

We shall choose $v$=$0$, corresponding to a variant of minimal
subtraction, a choice which one can motivate by the observed dominance
of the leading-$b$ term in $\MS$ noted in section 2.
For reasons of simplicity we shall choose the 't Hooft scheme,
corresponding to $B(x)$=$1+cx$. With these choices one has
\be
F(\infty)=b\ln\frac{Q}{\Lambda_{\MS}}-\frac{5}{6}b\;;
\ee
and we shall use this in the resummations. It corresponds simply to
taking `$a$' as the one-loop coupling in the $\MS$
scheme with $\mu$=$\mbox{e}^{-5/6}Q$ (the V-scheme):
\be
a_{\mbox{\scriptsize 1-loop \normalsize}}=
\frac{1}{b\ln\frac{Q}{\Lambda_{V}}}
\;,
\ee
where $\Lambda_{V}$=$\mbox{e}^{5/6}\Lambda_{\MS}$. This is in
fact the same choice for $a$ as in references [1--3], where it is
motivated by noting that using the one-loop form for $a(\mu^{2})$
makes the leading-$b$ summation $\mu$-independent. We stress once
again that in our view its significance is that it maximizes
$D^{(L)}(a)$ for a given choice of finite scheme, $B(x)$, and
parameter $v$. In Figure~1 we show $\tilde{D}^{(L)}(a)$ plotted
versus `$a$' ('t Hooft scheme with $v$=$0$).
In the figures we have plotted versus $\tilde{a}=1-e^{-a}$, so that
the full RS variation can be fitted in a unit interval in $\tilde{a}$.
We have taken
$Q$=$M_{Z}$=$91\mbox{GeV}$ and
$\Lambda_{\MS}(N_{f}$=$5)$=$111\mbox{MeV}$. The solid curve
gives the overall $\tilde{D}^{(L)}(a)$, split into
$\tilde{D}^{(L)}(a)|_{UV}$ (dashed) and $\tilde{D}^{(L)}(a)|_{IR}$
(dashed-dot) contributions. Similar curves for $\tilde{K}^{(L)}(a)$
with $Q^{2}$=$2.5\mbox{GeV}^{2}$ and
$\Lambda_{\MS}(N_{f}$=$3)$=$201\mbox{MeV}$ are given in
Figure~2; and for $\tilde{R}^{(L)}(a)$ with $Q$=$91\mbox{GeV}$ in
Figure~3. The corresponding curve for $\tilde{R}^{(L)}_{\tau}(a)$ is
given in Figure~4, $Q$=$M_{\tau}$=$1.78\mbox{GeV}$ and
$\Lambda_{\MS}(N_{f}$=$3)$=$201\mbox{MeV}$. The qualitative
behaviour is as we described earlier. The relative sizes of the $UV$
and $IR$ contributions reflect the disposition of the $UV$ and $IR$
singularities described above for the different quantities.

\begin{figure}
\begin{center}
{}~\epsfig{file=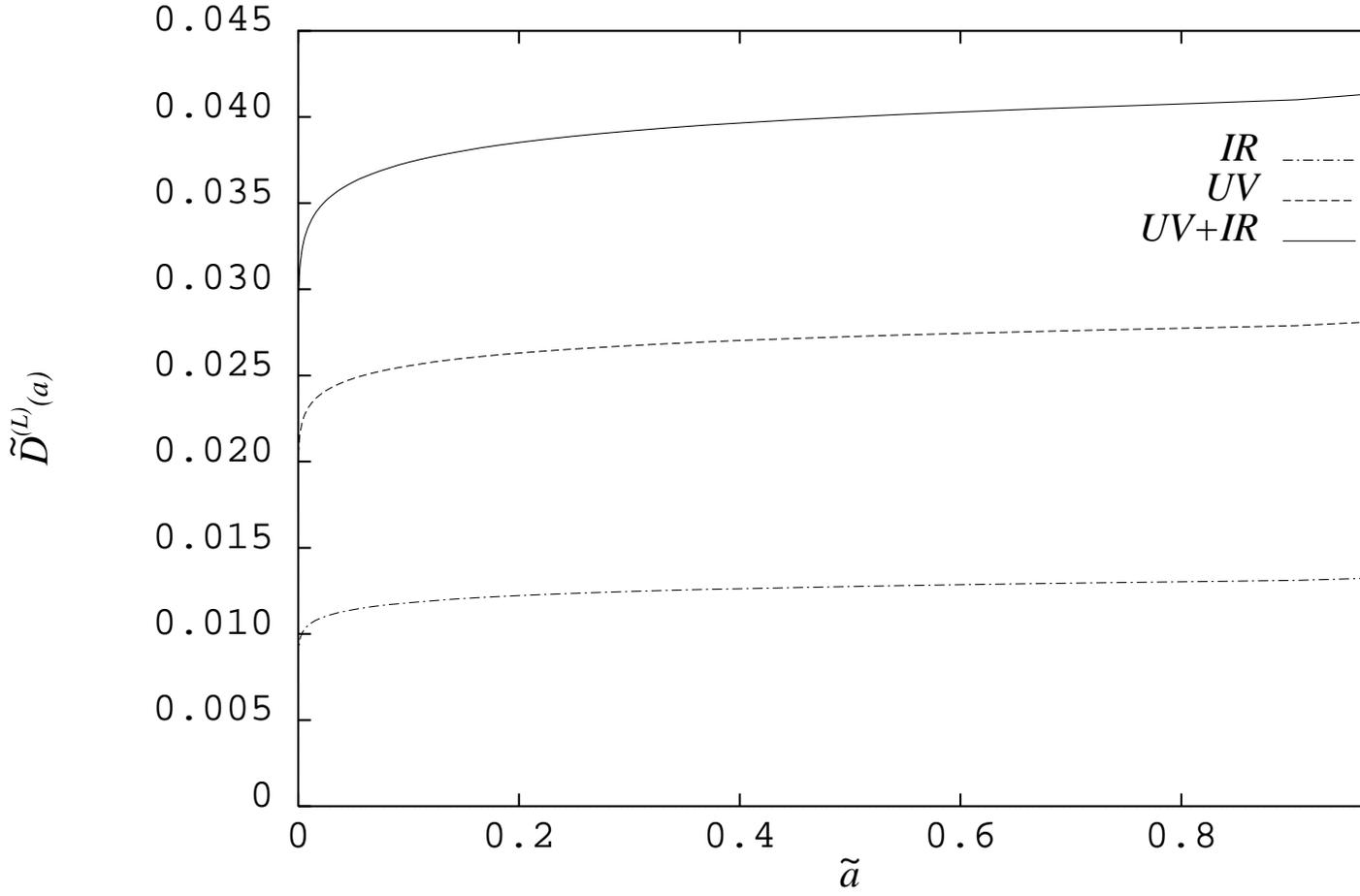,width=5.5in,angle=-90}
\caption{The leading-$b$ resummation,
$\tilde{D}^{(L)}(a)$, plotted versus $\tilde{a}$=$1-\mbox{e}^{-a}$
('t Hooft scheme $v$=$0$), $Q$=$M_{Z}$=$91\mbox{GeV}$ and
$\Lambda_{\MS}(N_{f}$=$5)$ is as in the text. The solid curve is
the overall result split into
$\tilde{D}^{(L)}(a)|_{UV}$ (dashed) and $\tilde{D}^{(L)}(a)|_{IR}$
(dashed-dot) contributions.}
\end{center}
\end{figure}
\begin{figure}
\begin{center}
{}~\epsfig{file=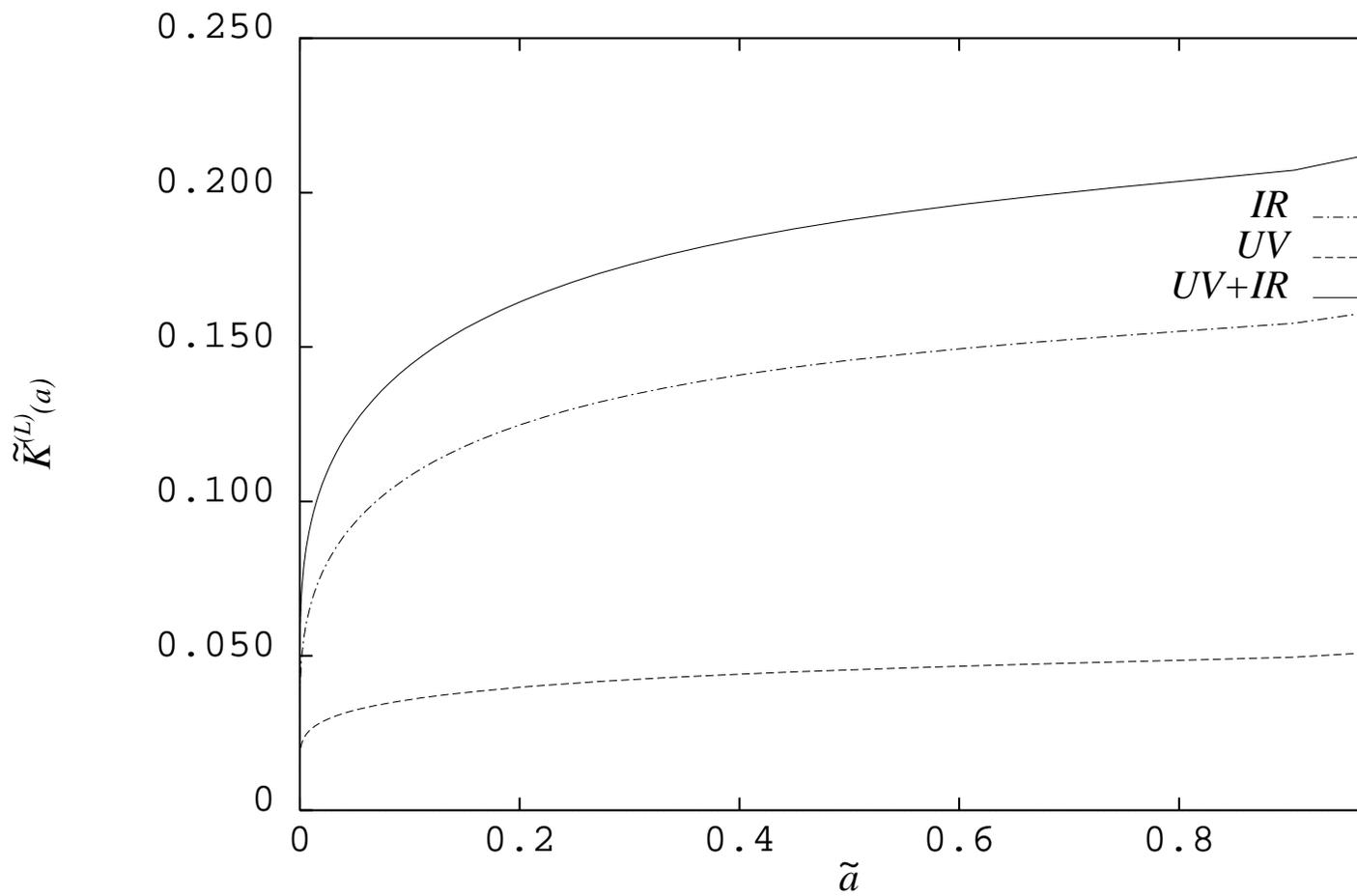,width=5.5in,angle=-90}
\caption{As for Figure~1 but for $\tilde{K}^{(L)}(a)$ with
$Q^{2}$=$2.5\mbox{GeV}^{2}$.}
\end{center}
\end{figure}
\newpage
\begin{figure}
\begin{center}
{}~\epsfig{file=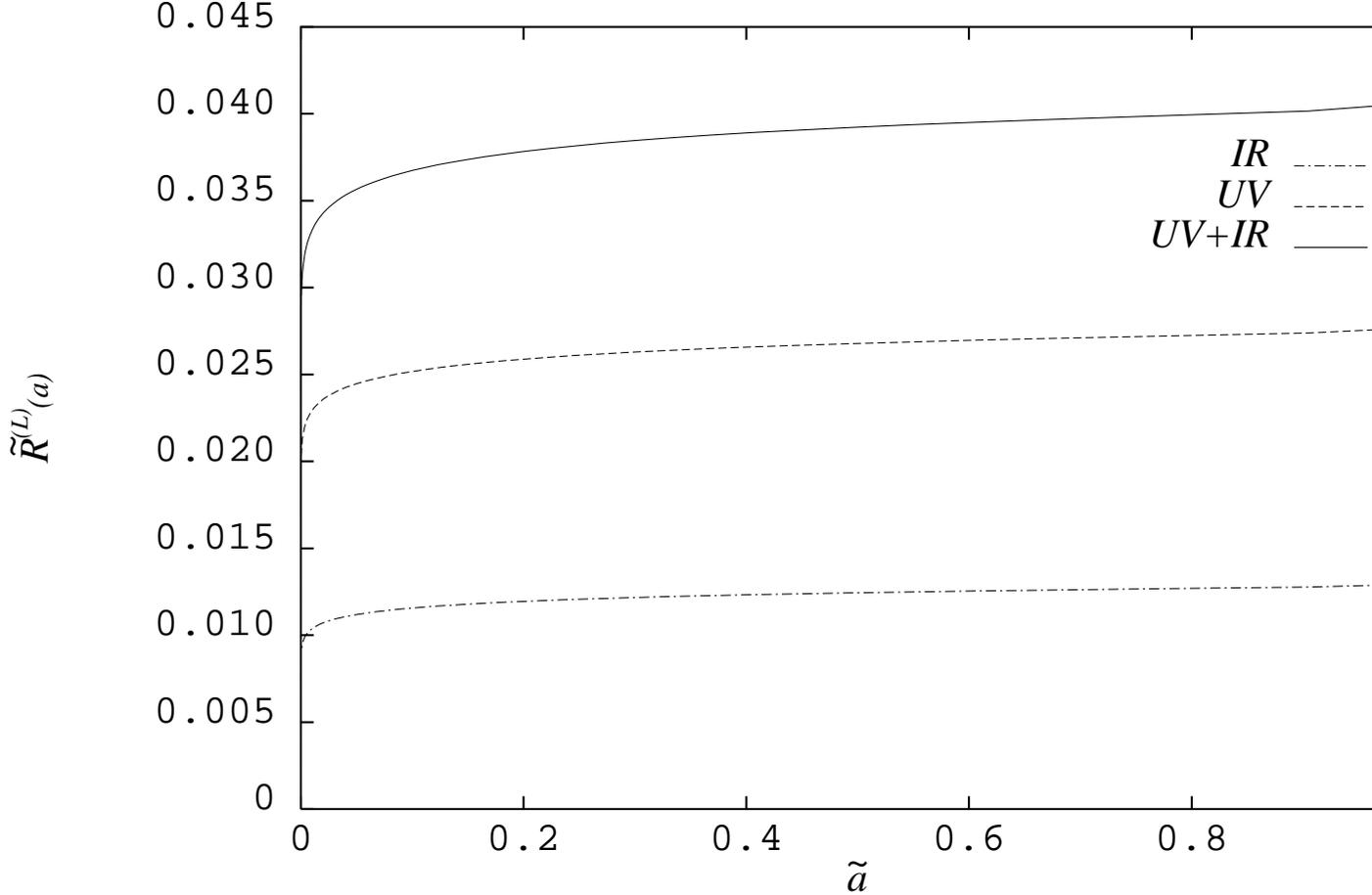,width=5.5in,angle=-90}
\caption{As for Figure~1 but for $\tilde{R}^{(L)}(a)$ with
$Q$=$M_{Z}$=$91\mbox{GeV}$.}
\end{center}
\end{figure}
\begin{figure}
\begin{center}
{}~\epsfig{file=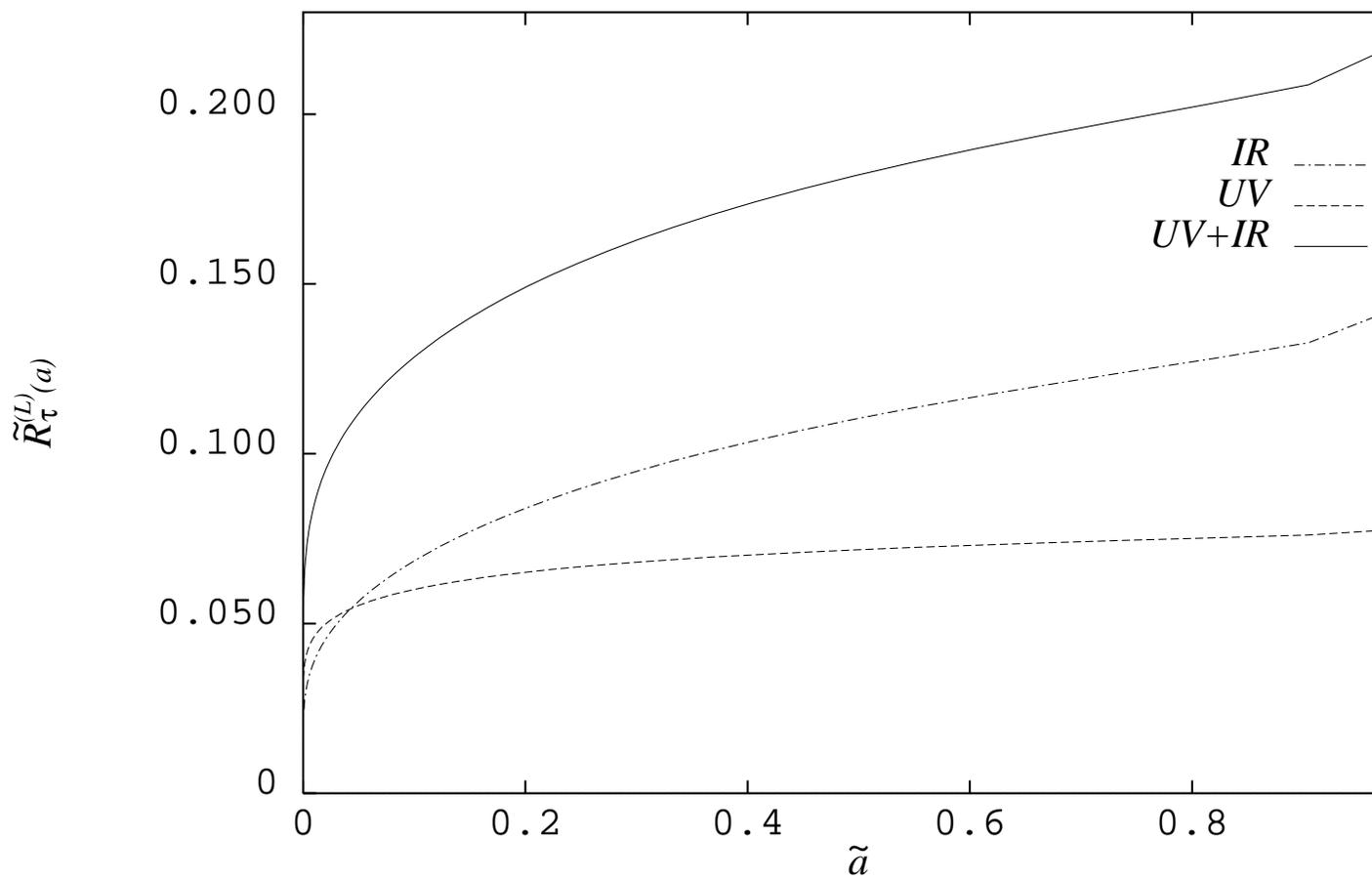,width=5.5in,angle=-90}
\caption{As for Figure~1 but for $\tilde{R}_{\tau}^{(L)}(a)$ with
$Q$=$M_{\tau}$=$1.78\mbox{GeV}$.}
\end{center}
\end{figure}

\begin{table}
\begin{center}
\begin{tabular}{|c|c|c|c|c|}\hline
\multicolumn{1}{|c|}{Observable}
&\multicolumn{1}{c|}{Energy}
&\multicolumn{1}{c|}{Resummed}
&\multicolumn{1}{c|}{FOPT}
&\multicolumn{1}{c|}{Expt}\\
\hline
$D$&$m_{\tau}$&1.151&1.087&---\\
&$Q_{0}$&1.055&1.045&---\\
&$m_{Z}$&1.042&1.035&---\\
\hline
$R$&$m_{\tau}$&1.105&1.080&---\\
&$Q_{0}$&1.053&1.044&---\\
&$m_{Z}$&1.041&1.035&$1.040\pm 0.004$\\
\hline
$R_{\tau}$&$m_{\tau}$&1.228&1.115&$1.183\pm 0.010$\\
\hline
$K$&$Q_{1}$&.784&.889&$.768\pm 0.09$\\\hline
\end{tabular}
\caption[t]{Comparison of resummed results
(`Resummed') of section~3 using $F(\infty)$, equation (74) (see text).
$Q_{0}^{2}$=$(20\mbox{GeV})^{2}$, $Q_{1}^{2}$=$2.5\mbox{GeV}^{2}$.
`FOPT' gives the exact NNLO
perturbative results with $\mu$=$Q$ $\MS$ scheme and the
$\Lambda_{\MS}$ values noted in the text. `Expt' gives the
experimentally deduced values for some of these quantities [25--27].}
\end{center}
\end{table}
The $a\rightarrow\infty$ limits obtained using $F(\infty)$ as in
equation (74) are tabulated in Table 3. Specifically, the `Resummed'
column contains $1+\tilde{D}^{(L)}(\infty)$,
$1+\tilde{R}^{(L)}(\infty)$, $1+\tilde{R}_{\tau}^{(L)}(\infty)$ and
$1-\tilde{K}^{(L)}(\infty)$ for each observable and we have added some
extra energies, $Q_{0}$=$20\mbox{GeV}$ and
$Q_{1}^{2}$=$2.5\mbox{GeV}^{2}$. The values of
$\Lambda_{\MS}$ for $N_{f}$=$5,3$
are as noted above. In each case we have taken care to
include
sufficient terms in the
summation over $UV_{\ell}$ and $IR_{\ell}$ singularities to guarantee
accuracy to the quoted number of significant figures.

As noted, using $F(\infty)$ is equivalent to the one-loop definition
of the coupling used in references [1--3]; and the same value of
$\Lambda_{\MS}(N_{f}$=$3)$ has been used. Our resummed results for
$D(m_{\tau}^{2})$, $R(m_{\tau}^{2})$ agree with the principal value
of the Borel sum for these quantities quoted in reference
\cite{neubert}; and the resummed result for $R_{\tau}$ agrees with the
results quoted in references \cite{ballbenbraun,neubert}. As discussed
at the end of section~3 we can reproduce the results of procedures `1'
and `2' for $R$ and $R_{\tau}$ in reference \cite{neubert} by
incorrectly omitting the second term in equation~(63) for some of the
$IR$ renormalon singularities.

There is clearly an ambiguity associated with the $IR$ renormalons or,
correspondingly, with vacuum condensates in the OPE. For a leading
$IR_{\ell}$ singularity in which there is a single pole one would
expect an ambiguity in the principal value
$\sim\tilde{B}\,\mbox{e}^{-z_{\ell}/a}$, where $\tilde{B}$ is the
residue
of the renormalon. For $\tilde{D}$ the leading $IR$ singularity is
$IR_{2}$. For the
\setlength{\leftmargin}{-.4in}
Minkowski quantities, $\tilde{R}$ and
$\tilde{R}_{\tau}$, the $\frac{\sin(\pi bz/2)}{\pi bz/2}$ factor
apparently removes the single pole at
$z$=$z_{2}$ but there is
presumably still a branch point singularity at $z$=$z_{2}$ beyond the
leading-$N_{f}$ approximation \cite{beneke}. The determination of the
residue
$\tilde{B}$ would require a resummation of the numerator polynomial
and so it is only known to leading-$N_{f}$ (for $\tilde{D}$
$B_{0}(2)$=$1$). Taking $\tilde{B}$=$1$ and putting $N_{f}$=$3,5$
values
for $b$ and `$a$' values corresponding to the energies and
$\Lambda_{\MS}$ considered in Table~3 yields an ambiguity
\raisebox{-.7ex}{$\stackrel{\textstyle <}{\sim}$}$10^{-4}$
for $\tilde{D}$, $\tilde{R}$, $\tilde{R}_{\tau}$, so the
significant figures quoted in the resummed result do not change. For
$\tilde{K}$, however, the leading singularity is $IR_{1}$ and one
would estimate the $IR$ ambiguity $\sim10^{-2}$ for $N_{f}$=$3$ and
$Q^{2}$=$2.5\mbox{GeV}^{2}$, which is clearly significant.
\setlength{\leftmargin}{0in}

The column labelled `FOPT' gives the fixed order perturbation theory
results obtained at NNLO (up to $O(a^{3})$) using the exact
perturbative
coefficients in the $\MS$ scheme with $\mu$=$Q$
[12--15].
The coupling `$a$' is defined using the NNLO truncated beta-function,
$B(x)$=$1+cx+c_{2}^{\overline{{\rm MS}}}x^{2}$, in equation (33), with
the values of $\Lambda_{\MS}$ as above.

The column labelled `Expt' gives the values determined from
experimental data for $(1$+$\tilde{R})$ \cite{siggi},
$(1$+$\tilde{R}_{\tau})$ \cite{pich}
and $(1$--$\tilde{K})$ \cite{ellis}.
The results of adjusting $\Lambda_{\MS}$ to fit
the
FOPT and resummed predictions to these experimental values
are summarised in Table~4.
\begin{table}
\begin{center}
\begin{tabular}{|c|c|c|c|}\hline
\multicolumn{1}{|c|}{\raisebox{-10pt}{Observable}}
&\multicolumn{1}{c|}{\raisebox{-10pt}{$N_{f}$}}
&\multicolumn{2}{c|}{\raisebox{-10pt}{$\Lambda_{\MS}$/MeV
fitted to experimental data}}\\
\multicolumn{1}{|c|}{ }
&\multicolumn{1}{c|}{ }
&\multicolumn{2}{c|}{ }\\
\cline{3-4}
& &\multicolumn{1}{|c|}{\raisebox{-10pt}
{\hbox{\hspace{18pt}}Resummed\hbox{\hspace{18pt}}}}
&\multicolumn{1}{c|}{\raisebox{-10pt}{NNLO FOPT}}\\
& &\multicolumn{1}{|c|}{ }
&\multicolumn{1}{c|}{ }
\\\hline
\raisebox{-10pt}{$R$}
&\raisebox{-10pt}{5}
&\raisebox{-10pt}{$98^{+93}_{-55}$}
&\raisebox{-10pt}{$287^{+226}_{-145}$}\\
&&&\\\hline
\raisebox{-10pt}{$R_{\tau}$}
&\raisebox{-10pt}{3}
&\raisebox{-10pt}{$159^{+9}_{-10}$}
&\raisebox{-10pt}{$386^{+21}_{-26}$}\\
&&&\\\hline
\raisebox{-10pt}{$K$}
&\raisebox{-10pt}{3}
&\raisebox{-10pt}{$375^{+20}_{-33}$}
&\raisebox{-10pt}{$495^{+15}_{-16}$}\\
&&&\\\hline
\end{tabular}
\caption[t]{Values of $\Lambda_{\MS}$ adjusted to fit the predictions
of NNLO FOPT and the resummed results
to the experimental data for $(1$+$\tilde{R})$ \cite{siggi},
$(1$+$\tilde{R}_{\tau})$ \cite{pich}
and $(1$--$\tilde{K})$ \cite{ellis}.}
\end{center}
\end{table}
Both fixed order perturbation theory and the leading-$b$ resummed
results exhibit RS-dependence and it is not at all obvious which
procedure gives the closest approximation to the all-orders sum. We
will defer a more detailed discussion of this question, including a
consideration of how the resummed results compare with use of the
effective charge formalism \cite{grun2,matt}, until a future work.

To conclude this section let us consider how the leading-$b$
resummation
might be improved. An obvious improvement would be to include the full
branch point structure of the renormalon singularities by
incorporating
the subleading in $N_{f}$, $\overline{\alpha}_{\ell}$ and
$\overline{\beta}_{\ell}$, pieces of the exponents
in equation (37) into the resummations. One expects
\ba
\overline{\alpha}_{\ell}&=&-cz_{\ell}+\gamma_{\ell}\;,\nonumber\\
\overline{\beta}_{\ell}&=&cz_{\ell}+\gamma_{\ell}'\;.
\ea
The first term can be deduced from RG considerations but the
$\gamma_{\ell}$ and $\gamma_{\ell}'$ are the one-loop anomalous
dimensions of the
relevant operators \cite{grun,vainzak}. For $\tilde{D}$ it is
known that $IR_{2}$ has a
corresponding OPE operator with vanishing one-loop anomalous
dimension,
$\gamma_{2}'$=$0$ \cite{al}, so $\overline{\beta}_{2}$=$cz_{2}$ is
known. To the
best of our knowledge the remaining
$\gamma_{\ell}$ and $\gamma_{\ell}'$'s are not known. One could
nonetheless include the first RG-predictable terms in equation (76) in
the resummations and see by how much the results change. The
problematic RS-dependence of $D^{(L)}(a)$ would be qualitatively
unchanged, however.

Further improvement could be achieved by including some of the
subleading in $N_{f}$, barred, coefficients in the numerator
polynomials
in equation (37). A complete fixed order perturbative calculation for
$D$ up to $O(a^{n})$ would enable the series coefficients of $B[D](z)$
up to $O(z^{n})$ to be determined exactly; but to obtain the
coefficients of the numerator polynomials for each singularity up to
$O(z^{n})$, even given knowledge of the full branch point exponents
discussed above, would still be very difficult. The improvement of
perturbation series by developing a representation of the form of
equation (37) with truncated numerator polynomials was suggested in
reference \cite{al}.

One further caveat concerns the structure of the $UV_{\ell}$
singularities. It has been suggested in reference
\cite{vainzak} that diagrams with
more than one renormalon chain will modify the form of the ultraviolet
renormalon singularities \cite{us};
this would lead to additional $UV$
terms in equation (37) with $O(z)$ or higher, subleading in $N_{f}$,
numerator coefficients but with increasing leading-$N_{f}$ exponents
$\alpha_{\ell}$. The presence of such terms would destroy the
asymptotic dominance of the leading-$b$ coefficient to all-orders in
$N_{f}$ \cite{us},
which seemed to be already evident in the comparisons with
the exact NLO and NNLO coefficients discussed in section~2. The
motivation for leading-$b$ resummation would hence disappear if this
$UV$ structure is correct. Further clarification of this point is
obviously required.

\section{Conclusions}

In this paper we have investigated the possibility of resummation to
all
orders of the leading term in the `$b$-expansion' of QCD perturbative
coefficients in equation (3). This expansion was introduced and
motivated in reference \cite{us} by a consideration of renormalon
singularities
in the Borel plane; and, if such singularities are present, then the
$d_{k}^{(k)}b^{k}$ term should, when expanded in $N_{f}$, reproduce
the
$N_{f}$-expansion coefficients of equation (2) to all orders in
$N_{f}$
with asymptotic accuracy $O(1/k)$. We checked explicitly in section~2
that for the QCD Adler $D$-function ($\tilde{D}$) and Deep Inelastic
Scattering sum rules ($\tilde{K}$) this asymptotic dominance of the
leading-$b$ terms was already evident in comparisons with the exact
NLO
and NNLO perturbative coefficients for those quantities. The
interesting
absence of $\zeta$-functions from the $d_{k}^{(0)}$ coefficient which
survives in the $b\rightarrow0$ limit was also noted. The RS
dependence
of the leading-$b$ coefficient and the need to give an all-orders
definition of the coupling in order to perform resummations was
discussed.

In section 3 we used exact large-$N_{f}$ results [4--6] to obtain
partial
information about the Borel transforms $B[\tilde{D}](z)$ and
$B[\tilde{K}](z)$. Ultra-violet and infra-red renormalon singularities
are present and we obtained the constant coefficients in the numerator
polynomials exactly; and the exponents, single and double poles, to
leading-$N_{f}$ (equations (40), (43)). We showed that in the V-scheme
($\MS$ with $\mu$=$\mbox{e}^{-5/6}Q$) the leading-$b$
coefficients, $d_{n}^{(n)}$ and $K_{n}^{(n)}$, are dominated,
even in low
orders, by the renormalon singularities nearest the origin,
respectively
$UV_{1}$ and $UV_{1}$+$IR_{1}$ combined (see Tables 1 \& 2).

For each quantity we split the leading-$b$ Borel sum into $UV$ and
$IR$
poles. The first contribution could be evaluated exactly in terms of
the
exponential integral function Ei($x$) (equation (47)) and elementary
functions; and the second, $IR$, contribution could be obtained as a
principal value, in terms of a principal value of Ei($x$). We showed
how to modify the Borel transform for the Minkowski continuations of
$\tilde{D}$, the $e^{+}e^{-}$ $R$-ratio ($\tilde{R}$) and the
$R$-ratio
for $\tau$-decay  ($\tilde{R}_{\tau}$), and a similar resummation was
performed
for these Minkowski quantities in terms of a generalised Ei($n,w$)
function (equations (58) and (62)).

In this way we obtained the $D^{(L)}$ component of the split defined
in
equation (4) for the above quantities. Unfortunately the result
obtained
by summing the leading-$b$ terms is RS-dependent, $D^{(L)}(a)$. In
section~4 we showed that one can maximize $D^{(L)}(a)$ and hence
perhaps
minimize $D^{(NL)}(a)$ for any given choice of finite scheme and
subtraction procedure. Maximizing $D^{(L)}(a)$ whilst choosing the
't Hooft scheme and a variant of minimal subtraction ($v$=$0$) was
equivalent to using the one-loop coupling in the V-scheme, the choice
also made in references [1--3].

We compared our resummed results with the principal values of the
Borel sum for $D(m_{\tau}^{2})$ and $R(m_{\tau}^{2})$ quoted in
reference \cite{neubert} and with that for $R_{\tau}$ in references
\cite{ballbenbraun,neubert} and found agreement. They are tabulated in
Table~3. The procedures `1' and `2' for continuing the running
coupling representation of reference \cite{neubert}, for $R$ and
$R_{\tau}$, to the Minkowski region were shown to correspond to using
different continuations of the Ei($n,w$) function from Re$\,w>0$ to
Re$\,w<0$. Only a continuation of Ei($n,w$) as a real function enables
one to evaluate correctly the well-defined pieces of the Borel sum
involving single $IR$ renormalon poles with a $\sin\frac{\pi bz}{2}$
factor. The inequivalent procedures of reference \cite{neubert} are
seen to correspond to various ways of wrongly evaluating the Borel sum
and are therefore spurious.
We stress that the Borel sum with $IR$
singularities identified and principal value regulated provides a
unique
result and, in our view, a firm foundation for combining vacuum
condensates in the OPE with $IR$ renormalons to achieve a well-defined
overall result (see reference \cite{grun}).
There is, of course, ambiguity due to
the $IR$ poles and we estimated this to be
\raisebox{-.7ex}{$\stackrel{\textstyle <}{\sim}$}$10^{-4}$ for the
$\tilde{D}$, $\tilde{R}$, $\tilde{R}_{\tau}$ resummed results in
Table~3, so the displayed significant figures should be valid; but
much
larger, $\sim10^{-2}$, for $\tilde{K}$ due to the presence of an
$IR_{1}$ singularity.

We also compared with fixed order perturbation theory up to NNLO for
these quantities. Since both $D^{(L)}(a)$ and the NNLO fixed order
result
suffer from RS dependence it is unclear which is the more reliable and
further investigation of this question is required.

We finally considered how the leading-$b$ resummation might be
improved
by including more exact information about the Borel transform.

\section*{Acknowledgements}

We would like to thank David Broadhurst for a number of stimulating
discussions and Martin Beneke for useful correspondence about the
curious connections between $IR$ and $UV$ renormalon residues for
vacuum polarization, noted in section~2.

Matthias Neubert and Martin Beneke are thanked for pointing out an
error affecting the resummations for the Minkowski quantities in an
earlier version of this paper.

CL-T gratefully acknowledges receipt of a P.P.A.R.C. U.K. Studentship.


\newpage
\begin{thebibliography}{99}
\bibitem{benbraun1} M.Beneke and V.M.Braun, DESY--94--200,
[hep--ph/9411229].
\bibitem{ballbenbraun} P.Ball, M.Beneke and V.M.Braun,
CERN--TH/95--26, [hep--ph/9502300].
\bibitem{neubert} M.Neubert, CERN--TH7524/94, [hep--ph/9502264].
\bibitem{beneke} M.Beneke, Nucl. Phys. {\bf B405} (1993) 424.
\bibitem{broad} D.J.Broadhurst, Z. Phys. {\bf C58} (1993) 339.
\bibitem{broadkat} D.J.Broadhurst and A.L.Kataev, Phys. Lett.
{\bf B315} (1993) 179.
\bibitem{benbraun2} M.Beneke and V.M.Braun, Nucl. Phys. {\bf B426}
(1994) 301.
\bibitem{us} C.N.Lovett--Turner and C.J.Maxwell, Nucl. Phys.
{\bf B432} (1994) 147.
\bibitem{grun} G.Grunberg, Phys. Lett. {\bf B325} (1994) 441.
\bibitem{blm} S.J.Brodsky, G.P.Lepage and P.B.Mackenzie, Phys. Rev.
{\bf D28} (1983) 228.
\bibitem{broadgroz} D.J.Broadhurst and A.G.Grozin, OUT--4102--52,
[hep--ph/9410240].
\bibitem{oldrussians1} K.G.Chetyrkin, A.L.Kataev and F.V.Tkachov,
Phys. Lett. {\bf B85} (1979) 277;\newline
M.Dine and J.Sapirstein, Phys. Rev. Lett.
{\bf 43} (1979) 668; \newline
W.Celmaster and R.J.Gonsalves, Phys. Rev. Lett.
{\bf 44} (1980) 560.
\bibitem{newrussians1} S.G.Gorishny, A.L.Kataev and S.A.Larin, Phys.
Lett.
{\bf B259} (1991) 144;\newline
L.R.Surguladze and M.A.Samuel, Phys. Rev. Lett. {\bf 66} (1991) 560;
{\bf 66} (1991) 2416 (Erratum).
\bibitem{newrussians2} S.G.Gorishny and S.A.Larin, Phys. Lett.
{\bf B172} (1986) 109; \newline
E.B.Zijlstra and W. van Neerven, Phys. Lett.
{\bf B297} (1992) 377.
\bibitem{verma} S.A.Larin and J.A.M.Vermaseren, Phys. Lett. {\bf B259}
(1991) 345.
\bibitem{brodlu} S.J.Brodsky and H.J.Lu, Phys. Rev. {\bf D51}
(1995) 3652.
\bibitem{origoa3} S.G.Gorishny, A.L.Kataev and S.A.Larin, Phys.
Lett. {\bf B212} (1988) 238.
\bibitem{gbstrs} W.Celmaster and R.J.Gonsalves, Phys. Rev. {\bf D20}
(1979) 1420.
\bibitem{vold} D.R.T.Jones, Nucl. Phys. {\bf B75} (1974) 537;\newline
W.E.Caswell, Phys. Lett. {\bf 33} (1974) 244.
\bibitem{spew} P.M.Stevenson, Phys. Rev. {\bf D23} (1981) 2916.
\bibitem{thooft} G.'t Hooft in ``The Whys of Subnuclear Physics'',
Erice 1977, ed. A.Zichichi (Plenum, New York 1977).
\bibitem{benthes} M.Beneke, Ph.D. Thesis ``Die Struktur der
St\"{o}rungsreihe in hohen Ordnungen'', M\"{u}nchen, 1993.
\bibitem{gg} X.Ji, MIT--CTP--2381 (1994), [hep--ph/9411312].
\bibitem{bnp} E.Braaten, S.Narison and A.Pich, Nucl. Phys. {\bf B373}
(1992) 581.
\bibitem{siggi} LEP Collaborations Joint Report No.
CERN-PPE/93-157 (1993) .
\bibitem{pich} This value is quoted in the review by A.Pich, preprint
FTUV--94--71 (1994), [hep--ph/9412273].
\bibitem{ellis} J.Ellis and M.Karliner, Phys. Lett. {\bf B341} (1995)
397.
\bibitem{grun2} G.Grunberg, Phys. Lett. {\bf B95} (1980) 70;\newline
G.Grunberg, Phys. Rev. {\bf D29} (1984) 2315.
\bibitem{matt} D.T.Barclay, C.J.Maxwell and M.T.Reader, Phys. Rev.
{\bf D49} (1994) 3480.
\bibitem{vainzak} A.I.Vainshtein and V.I.Zakharov,  Phys. Rev. Lett.
{\bf 73} (1994) 1207.
\bibitem{al} A.H.Mueller, Nucl. Phys. {\bf B250} (1985) 327.
\bibitem{abst} Handbook of Mathematical Functions, ed. Milton
Abramowitz and Irene A.Stegan (Dover), Section 5.1, p228.
\end{thebibliography}
\end{document}